\documentstyle[12pt,epsf]{article}
\def\journal#1, #2, #3, #4 { {\sl #1~}{\bf #2~}(#3) #4 }

\def\prd{\journal Phys. Rev. D, }

\def\prl{\journal Phys. Rev. Lett., }

\def\ijmp{\journal Int. J. Mod. Phys., }

\def\cmp{\journal Comm. Math. Phys., }

\def\np{\journal Nucl. Phys., }

\def\pl{\journal Phys. Lett., }

\def\annp{\journal Ann. Phys. (N.Y.), }

\catcode`\@=11
\def\marginnote#1{}
\newcount\hour
\newcount\minute
\newtoks\amorpm
\hour=\time\divide\hour by60
\minute=\time{\multiply\hour by60 \global\advance\minute
by-\hour}\edef\standardtime{{\ifnum\hour<12
\global\amorpm={am}%
        \else\global\amorpm={pm}\advance\hour by-12 \fi
        \ifnum\hour=0 \hour=12 \fi
        \number\hour:\ifnum\minute<10
0\fi\number\minute\the\amorpm}}
\edef\militarytime{\number\hour:\ifnum\minute<10
0\fi\number\minute}

\def\draftlabel#1{{\@bsphack\if@filesw {\let\thepage\relax
   \xdef\@gtempa{\write\@auxout{\string
      \newlabel{#1}{{\@currentlabel}{\thepage}}}}}\@gtempa
   \if@nobreak \ifvmode\nobreak\fi\fi\fi\@esphack}
        \gdef\@eqnlabel{#1}}
\def\@eqnlabel{}
\def\@vacuum{}
\def\draftmarginnote#1{\marginpar{\raggedright\scriptsize\tt#1}}
\def\draft{\oddsidemargin -.5truein
        \def\@oddfoot{\sl preliminary draft \hfil
        \rm\thepage\hfil\sl\today\quad\militarytime}
        \let\@evenfoot\@oddfoot \overfullrule 3pt
        \let\label=\draftlabel
        \let\marginnote=\draftmarginnote

\def\@eqnnum{(\theequation)\rlap{\kern\marginparsep\tt\@eqnlabel}%
\global\let\@eqnlabel\@vacuum}  }

\def\numberbysection{\@addtoreset{equation}{section}
        \def\theequation{\thesection.\arabic{equation}}}

\def\underline#1{\relax\ifmmode\@@underline#1\else
 $\@@underline{\hbox{#1}}$\relax\fi}

\catcode`@=12
\relax

\numberbysection
\def\souligne#1{\underline{#1}}
\def\fin{\end{document}}
\def\beq{\begin{equation}}
\def\eeq{\end{equation}}
\def\beqa{\begin{eqnarray}}
\def\eeqa{\end{eqnarray}}
 \def\nnn{\nonumber \\}
\def\sqr#1#2{{\vcenter{\vbox{\hrule height.#2pt
\hbox{\vrule width.#2pt height#1pt \kern#1pt
\vrule width.#2pt}
\hrule height.#2pt}}}}

\def\Je{J^e}

\def\Jne#1 {J_{#1}^e\, \!}
\def\Jneb#1 {{\overline J}_{#1}^e\, \!}
\def\Jnep#1 {J_{#1}'\, \!^e\, \!}
\def\Jnebp#1 {{\overline J}_{#1}'\, \!  \!^e\, \!}

\def\Jehat{{\widehat J}^e}
\def\hhat{{\widehat h}}

\def\Jhat{{\widehat J}}

\def\mhat{{\widehat m}}

\def\Vhat{{\widehat V}}

\def\psihat{{\widehat \psi}}

\def\Vhat{{\widehat V}}

\def\qhat{{\widehat q}}

\def\varpihat{{\widehat \varpi}}

\def\varpib{{\overline \varpi}}

\def\mhat{{\widehat m}}

\def\phat{{\widehat p}}

\def\zb{{\bar z}}
\def\nb{{\bar n}}

\def\Vb{{\overline V}}

\def\Vt{{\widetilde V}}

\def\Jge{{\underline J}}
\def\mge{{\underline m}}

\def\Jgen#1 {  {\underline J_{#1}} }
\def\Jgenp#1 #2 {(J_{#1}+{#2},\Jhat_{#1})}
\def\Jgenm#1 #2 {(J_{#1}-{#2},\Jhat_{#1})}
\def\Jg#1 {J_{#1},\Jhat_{#1}}
\def\Jgp#1 #2 {J_{#1}+{#2},\Jhat_{#1}}
\def\Mgen#1 {{\underline M_{#1}}}

\def\produit#1,#2,#3,#4 {P\Bigl ( [{#1},{#2}]\otimes\{{#3}\},{#4}\Bigr )}
\def\produitscript#1,#2,#3,#4 {P\Bigl (
[{\scriptstyle{#1},{#2}}]\otimes\{{\scriptstyle{#3}}\},{#4}\Bigr
)}
\def\pprod#1,#2,#3,#4,#5 {P\Bigl ( [{#1},{#2}]\otimes[{#3},{#4}],{#5}\Bigr )}
\def\pprodscript#1,#2,#3,#4,#5 {P\Bigl (
[{\scriptstyle{#1},{#2}}]\otimes[{\scriptstyle{#3},{#4}}],{#5}\Bigr )}

\def\fusV#1,#2,#3,#4,#5,#6 {f_V(
\Jgen{#1} ,
\Jgen{#2} ,
\Jgen{#3} ,
\Jgen{#4} ,
\Jgen{#5} ,
\Jgen{#6} )}

\def\brdV#1,#2,#3,#4,#5,#6 {b_V(
\Jgen{#1} ,
\Jgen{#2} ,
\Jgen{#3} ,
\Jgen{#4} ,
\Jgen{#5} ,
\Jgen{#6} )}

\def\fusxi#1,#2,#3 {f_\xi (\Jgen{#1} ,
\Mgen{#1} ,
\Jgen{#2} ,
\Mgen{#2} ,
\Jgen{#3} )}

\def\gaghat{{\hat {\bigl \{}}}
\def\gadhat{{\hat {\bigr \}}}}

\def\bverthat{{\hat {\bigl \vert}}}

\def\sixjxi#1,#2,#3,#4,#5,#6 {{\left\{\left . \!\! \,^{#1}_{#2}
\,^{#3}_{#4} \right | \!\, ^{#5}_{#6}\right\}}}
\def\sixje#1,#2,#3,#4,#5,#6 {{\left\{\left\{\left . \!\! \, ^{#1}_{#2}
\, ^{#3}_{#4} \right | \!\, ^{#5}_{#6}\right\}\right\}}}
\def\sixjxihat#1,#2,#3,#4,#5,#6 {{{\gaghat\left . \!\! \, ^{#1}_{#2}
\, ^{#3}_{#4} \right | \!\, ^{#5}_{#6}\gadhat}}}
\def\sixjehat#1,#2,#3,#4,#5,#6 {{\gaghat\gaghat\left . \!\! \, ^{#1}_{#2}
\, ^{#3}_{#4} \right | \!\, ^{#5}_{#6}\gadhat\gadhat}}

\def\Jpe {J^{e+}}
\def\Jme {J^{e-}}

\def\Jehat {{\widehat J^e}}

\begin{document}
\begin{flushright}
hep-th/9408069
\end{flushright}

\vglue 2  true cm
\begin{center}
{\large \bf
Quantum deformations of $sl(2)$:         \\
\medskip
The hard core of
two dimensional gravity.           } \\
\vglue .5 true cm
{\bf Jean-Loup~GERVAIS}\\
\medskip
\medskip
{\footnotesize Laboratoire de Physique Th\'eorique de
l'\'Ecole Normale Sup\'erieure\footnote{Unit\'e Propre du
Centre National de la Recherche Scientifique,
associ\'ee \`a l'\'Ecole Normale Sup\'erieure et \`a
l'Universit\'e
de Paris-Sud.},\\
24 rue Lhomond, 75231 Paris CEDEX 05, ~France}.
\end{center}
\begin{abstract}
\baselineskip .4 true cm
\noindent
{\footnotesize
The quantum group structure of the Liouville theory is
reviewd and shown to be a important tool for
solving the theory.}
\end{abstract}
\section{Introduction}
The  Liouville theory arose more than ten years ago from Weyl anomaly
in two dimensions. It is actually the simplest member  of the
family of conformal Toda theories\cite{LS}  which should be considered as
W gravity in the conformal gauge\cite{BG,GM}. Upon quantization
quantum group structures emerge\cite{B,G,CG1,GS2}
such that the mathematical
parameters $h$ of the mathematical deformations coincide with
the Planck constants, with appropriate unit choices. Thus, the
non-commutativity which is inherent to the
group deformation -- since the co-product in non-symmetric --
 is brought about by
the very quantization of these systems. Such a
 quantum group structure was
already there from the beginning, in the early works of Neveu
and myself\cite{GN},
but in disguise. This
appearance of quantum group
seems  to be very natural geometrically, since one deals with a
gravity theory, where the space-time metric is quantized, which seems
tantamount to quantizing the two dimensional (2D)
 space-time itself. Thus an object like
a quantum plane should appear, and quantum groups
are natural transformations of such ``quantum manifolds''.
In these lecture notes, we shall not follow this last line, but
rather  review the basic features
of the Liouville theory, form the viewpoint
taken\cite{CGR1,CGR2,GS1,GS3,GR1}  recently
by E. Cremmer, J.-F. Roussel, J. Schnittger and myself, where the
operator algebra of the chiral components is completely determined
in terms of quantum group symbols, within the framework\cite{MS}  of
Moore and Seiberg. This result is instrumental in deriving the
detailed  properties of 2D gravity coupled with  2D matter.
 In the weak coupling regime of gravity,
this gives back\cite{G3} the results of matrix models in a way which may
seem unnecessarily painful. However, this quantum group approach
is the only one which extends to the strong coupling regime.
This recently allowed\cite{GR1}
 J.-F. Roussel and myself to actually solve
a new type of strongly coupled topological model, which we will
mention   at the end.

\section{Basic points about Liouville theory}
\subsection{The classical case}
In order to set the stage, we recall the classical structure
for the special case of the Liouville
theory. We shall work with Euclidean coordinates
$\sigma$, $\tau$. As a preparation for the quantum case, the
classical action is defined as
\begin{equation}
 {\it S}={1\over 8\pi } \int d\sigma d\tau
\, \Bigl ( ({\partial \Phi\over \partial \sigma})^2
+({\partial \Phi\over \partial \tau})^2
+e^{\displaystyle 2\sqrt{\gamma} \Phi} \Bigr ).
\label{c2.1}
\end{equation}
We use world sheet variables $\sigma$ and $\tau$, which are
local coordinates such
that the metric tensor takes the form $h_{ab} =\delta_{a,b}\,
e^{\displaystyle 2\Phi \sqrt{\gamma}}$. The
complex structure is assumed to be such that the curves  with
constant $\sigma$ and $\tau$ are everywhere
tangent to the local imaginary
and real axis  respectively.
The action \ref{c2.1}  corresponds to a conformal
 theory such that $\exp(2\sqrt{\gamma} \Phi) d\sigma d\tau $
is invariant. It is convenient to let
\beq
  x_{\pm }=\sigma\mp   i \tau ,
\quad
\partial_\pm = {1\over 2} ( \partial_\sigma \pm i \partial_\tau).
\label{2.2}
\eeq
By minimizing the above action, one derives the Liouville equations
\begin{equation}
 {\partial^2 \Phi\over \partial \sigma^2}
+{\partial^2 \Phi\over \partial \tau^2}=
 \sqrt{\gamma}\> e^{\displaystyle 2\sqrt{\gamma}\Phi}.
\label{2.3}
\end{equation}
The chiral modes
 may be separated very simply using the fact that
the field $\Phi (\sigma,\, \tau)$ satisfies the above equation
if and only if\footnote{ The factor $i$ means that these
solutions should be considered in Minkowsky space-time }
\beq
e^{-\displaystyle \sqrt{\gamma}\Phi}={i\sqrt{\gamma} \over 2}
\sum_{j=1,2} f_j(x_+)
g_j(x_-);
  \quad  x_\pm=\sigma\mp i\tau
\label{2.4}
\end{equation}
where $f_j$ (resp.($g_j$), which are functions of a single variable,  are
solutions of the  same Schr\"odinger equation (primes mean derivatives)
\beq
-f_j''+T(x_+)f_j=0,\quad
\hbox{( resp.}\>  -g''_j+ \overline T(x_-)g_j\,\hbox{)}.
\label{2.5}
\end{equation}
 The solutions
are normalized such that their Wronskians $f_1'f_2-f_1f_2'$
and $g_1'g_2-g_1g_2'$ are equal to one.
The proof goes as follows.

 1) First check that Eq.\ref{2.4} is indeed solution. Taking the Laplacian
 of the logarithm of the right-hand side gives
$${\partial^2 \Phi\over \partial \sigma^2}
+{\partial^2 \Phi\over \partial \tau^2}
\equiv 4\partial_+\partial_-\Phi
=-4 \sqrt{\gamma} \Bigl / \bigl( \sum_{i=1,2} f_i g_i\bigr)^2
$$
where $\partial_\pm=(\partial/\partial \sigma
\pm i\partial /\partial\tau)/2$  .
The numerator has been simplified by means of the Wronskian condition.
This is equivalent to Eq.\ref{2.3}.

2) Conversely check that any solution of Eq.\ref{2.3}
may be put
under the form  Eq.\ref{2.4}. If Eq.\ref{2.3} holds
one deduces
\beq
 \partial_\mp T^{(\pm)}=0;\quad \hbox{with}\,
T^{(\pm)}:=e^{\displaystyle \sqrt{\gamma} \Phi} \partial ^2_\pm
e^{-\displaystyle \sqrt{\gamma} \Phi}
\label{2.6}
\end{equation}
$T^{(\pm)}$ are thus functions of a single variable. Next
the equation involving $T^{(+)}$ may be rewritten as
\beq
(-\partial_+^2 +T^{(+)}\bigr)e^{-\displaystyle \Phi}=0
\label{2.7}
\end{equation}
with solution
$$e^{-\displaystyle \sqrt{\gamma} \Phi}=
{i\sqrt\gamma\over 2} \sum_{j=1,2} f_j(x_+) g_j(x_-);
\quad \hbox{with}\, -f_j''+T^{(+)}f_j=0
$$
where the $g_j$ are arbitrary functions of $x_-$. Using the equation
\ref{2.6} that involves $T^{(-)}$, one finally derives the
Schr\"odinger equation  $-g_j''+T^{(-)}g_j=0$. Thus the theorem holds
with $T=T^{(+)}$ and $\overline T=T^{(-)}$.
One may deduce from Eq.\ref{2.6} that  the potentials of the two Schr\"odinger
equations
coincide with the two chiral components of the stress-energy tensor.
 Thus these equations are the classical equivalent of the
Ward identities
that ensure the decoupling of Virasoro null vectors.
Next a B\"acklund transformation to free fields is easily obtained by
writing
\beq
f_1(x)=e^{q(x)\sqrt{\gamma}}, \quad f_2(x)=f_1(x)
\int^x dx_1 e^{-2q(x_1)\sqrt{\gamma}},
\label{2.8}
\end{equation}
It follows from the canonical Poisson brackets of Liouville theory that
\beq
 \Bigl \{ p(\sigma_1-i\tau),
p(\sigma_2-i \tau) \Bigr \}_{\hbox{PB}}=
2\pi   \delta'(\sigma_1-\sigma_2).
\label{a4.8}
\eeq
where $p=q'$.
Clearly the differential equation Eq.\ref{2.5}
gives
$T=p^2+p'/\sqrt{\gamma}$, or equivalently
$T=(q')^2+q''/\sqrt{\gamma}$. The last expression coincides with
$U_1$ Sugawara stress-energy tensor with a linear  term. From
the
viewpoint of differential equations, these relations are simply
the well-known Riccati equations associated with the Schr\"odinger
equation Eq.\ref{2.5}. An easy computation shows that
$T$ satisfies  the Poisson bracket Virasoro algebra with
$C_{\hbox{class}}=3/\gamma$. We shall consider the typical
situation of a
cylinder with $0\leq \sigma \leq \pi$, and $-\infty \leq \tau
\leq \infty$.  After appropriate coordinate
change, this may describe one handle of a surface with arbitrary genus. Then
$T$ is periodic in $\sigma$ with period $2\pi$, and it is
important to discuss the monodromy properties of the Schr\"odinger
equation, whose physics is that   of one-dimensional crystal.

Consider any pair $f_{1,2}$  of independent solutions of the
equations $-f''+Tf=0$. If $T(x+2\pi)=T(x)$, one has
$f_j(x+2\pi)=\sum_k m_{jk} f_k(x)$, where the monodromy matrix
$m_{jk}$  is
independent of $x$. In the generic case, the matrix is
diagonalizable,
and there exist the  so-called Bloch waves solutions
 $V_{1,2}$ such that  $V_j(x+2\pi)=\alpha_j  V_j(x)$. Since the
Wronskian of the two solution is a constant, the product of
eigenvalues $\alpha_1 \alpha_2$ is equal to one. We may identify
one of the two Bloch waves, say, $V_1$ with $f_1$ if we take
$p$ to be periodic. Indeed if we write
\[
p(x)=\sum_n e^{-inx} p_n
\]
we have
\[q(x)= q_0+p_0 x+i\sum_{n\not= 0} e^{-inx}  p_n/n
\]
and $V_1(x+2\pi) = \exp (2\pi p_0) V_1(x)$. The other Bloch wave
$V_2$ should be such that
$V_2(x+2\pi) = \exp (-2\pi p_0) V_2(x)$. This may  be easily
achieved by taking a linear combination of $f_1$ and $f_2$
in Eq.\ref{2.8}.
As a preparation for the quantum case, let us remark that, at
this
level, there is a complete symmetry between $V_1$ and $V_2$ which
are simply the two eigenvectors of the monodromy matrix (we
do not consider the so-called parabolic situation where  this
matrix would not be diagonolizable).  This is in
apparent contrast with the expression Eq.\ref{2.8} where $f_1$
played a special role. We may re-establish this symmetry by writing
\beq
V_i=e^{q^{[i]}(x)\sqrt{\gamma}}, \quad p^{[i]}= q^{[i]'},
\quad i=1,\, 2.
\label{a4.9}
\eeq
Since the $V$'s are solutions of the same Schr\"odinger equation,
the fields $p^{[i]}$  are related by the equation
\beq
T=(p^{[1]})^2+{p^{[1]'}\over \sqrt {\gamma}}=
(p^{[2]})^2+{p^{[2]'}\over \sqrt {\gamma}}
\label{a4.10}
\eeq
Thus the $p^{[i]}$'s
 are simply the two independent solutions of the Riccati
equations associated with our Schr\"odinger equation \ref{2.5}.
Of course $p^{[1]}$ coincides with the previous $p$ field, and thus
obey the P.B. algebra Eq.\ref{a4.8}. One may verify\cite{GN3}  that this
is
also true
for $p^{[2]}$. Thus we have
$$
 \Bigl \{ p^{[1]}(\sigma_1-i \tau),
p^{[1]}(\sigma_2-i  \tau) \Bigr \}_{\hbox{PB}}=
$$
\beq
\Bigl \{ p^{[2]}(\sigma_1-i  \tau),
p^{[2]}(\sigma_2-i  \tau) \Bigr \}_{\hbox{PB}}=
2\pi   \delta'(\sigma_1-\sigma_2),
\label{a4.11}
\eeq
 Clearly we may expand each field
in Fourier series:
\beq
q^{[j]}(x)=
q^{[j]}_0+p^{[j]}_0 x+i\sum_{n\not= 0} e^{-inx}  p^{[j]}_n/n.
\label{a4.12}
\eeq
Since the corresponding eigenvalues of the monodromy matrix have
their product equal to one, it follows that the $p^{[j]}_0$'s
are
related by
\beq
p^{[1]}_0=-p^{[2]}_0.
\label{a4.13}
\eeq
Eqs.\ref{a4.10}, and \ref{a4.13} define\cite{GN3}
 a canonical transformation
between two
equivalent free fields.
        This structure is  also present at the
quantum level.

Let return finally to the Liouville field Eq.\ref{2.4}. A priori,
any two
pairs
 of linearly independent solutions of Eq.\ref{2.5}
are suitable.  At this point it is convenient to introduce other
solutions than the Bloch waves  $V_j$, and $\bar V_j$ just
discussed. We hereafter call
$\overline f_i$ the  general pair describing the plus components. For the
minus components, it is convenient to change the definition so
that Eq.\ref{2.4} becomes
\begin{equation}
e^{-\displaystyle \sqrt{\gamma} \Phi}=
{i\sqrt{\gamma} \over 2}\left ( f_1(x_+)
\bar f_2(x_-)-f_2(x_+) \bar f_1(x_-)\right );
\label{ar2.4}
\end{equation}
With this convention, one  easily sees  that the Liouville exponential
 is
left unchanged if $f_j$ and $\overline f_j$ are replaced
by
$\sum_kM_{jk}f_k$ and   $\sum_kM_{kj}\overline f_k$,
respectively, where
$M_{jk}$ is an arbitrary constant matrix with
 determinant  equal to one.
Eq.\ref{ar2.4} is
$sl(2,C)$-invariant with $f_j$ transforming as a representation of
spin $1/2$. At the quantum level, the $f_j$'s and $\overline f_j$'s
will  become operators that do not commute, and  the
group $sl(2)$ will get  deformed to become  the quantum group
$U_q(sl(2))$. Note that the above Bloch-wave solutions
$V_i$ will not transform simply under action of the quantum group.
This is why we introduced other pairs  of solutions $f$, and $\bar
f$.
 This  structure plays a crucial role at
 the quantum level, and we now elaborate
upon  the classical  $sl(2)$ structure  where the calculations are
simple.
At the classical level, it is trivial to take Eq.\ref{ar2.4} to any
power.  For positive integer  powers $2J$, one gets
(letting $\beta=i{\sqrt {\gamma} \over 2}$)
$$
e^{-\displaystyle 2J \sqrt{\gamma} \Phi}=
$$
\beq
\sum_{M=-J}^J {\beta^{2J}(-1)^{J-M} (2J) !\over (J+M) !(J-M) !}
\left (f_1(x_+)\,\overline f_2(x_-)\right )^{J-M}
\left(f_2(x_+) \overline f_1(x_-)\right )^{J+M}.
\label{ar2.7}
\eeq
It is convenient  to put the result under the form
\begin{equation}
e^{-\displaystyle 2J \sqrt{\gamma} \Phi}=
\beta^{2J}
\sum_{M=-J}^J (-1)^{J-M}
f_M^{(J)}(x_+) \overline f_{-M}^{(J)}(x_-).
\label{ar2.8}
\end{equation}
where $J\pm M$ run over integer. The $sl(2)$-structure has been
made
transparent by letting
\beq
f_M^{(J)}\equiv \sqrt { \textstyle {2J\choose J+M}}
\left (f_1\right )^{J-M} \left(f_2\right )^{J+M},  \quad
\overline f_M^{(J)}\equiv\sqrt { \textstyle {2J\choose J+M}}
\left (\overline f_1\right )^{J+M}
\left(\overline f_2\right )^{J-M}.
\label{ar2.9}
\eeq
The notation anticipates that $f_M^{(J)}$ and $\overline
f_M^{(J)}$
form   representations of spin $J$. This is  indeed true
since
$f_1$, $f_2$ and $\overline f_1$, $\overline f_2$    span
spin $1/2$ representations, by construction.   Explicitely one finds
$$I_\pm f_M^{(J)}=\sqrt {(J\mp M)(J\pm M+1)} f_{M\pm 1}^{(J)},\quad
I_3 f_M^{(J)} =Mf_M^{(J)}
$$
\begin{equation}
\overline I_\pm \overline f_M^{(J)}
=\sqrt {(J\mp M)(J\pm M+1)} \overline f_{M\pm 1}^{(J)}, \quad
\overline I_3 f_M^{(J)} =M\overline f_M^{(J)},
\label{ar2.10}
\end{equation}
where $ I_\ell$ and $\overline I_\ell$ are the
infinitesimal generators of the   $x_+$ and  $x_-$ components
respectively. Moreover, one sees that
\begin{equation}
\left (I_\ell+ \overline I_\ell\right )
e^{-\displaystyle 2J \sqrt{\gamma} \Phi}=0
\label{ar2.11}
\end{equation}
so that the exponentials  of the Liouville field are group
invariants. For  Bloch waves, the corresponding chiral fields
are noted $V_m^{(J)}$, and $\Vb_m^{(J)}$. We shall mostly deal with them
in the quantum case. Finally, it is convenient to show
  how they  may be rewritten, at will,  in terms
of a single free field out of the two we have
introduced. Basically, one uses expression of the type \ref{2.8},
suitably modified to take account of the periodicity. It is
straightforward to verify that one may write
\beq
V^{(J)}_{-J}=({1\over \sqrt{S'}})^{2J}=
e^{ 2J\sqrt{\gamma}q^{[1]}}, \qquad
V^{(J)}_{J}=({S\over \sqrt{S'}})^{2J}=e^{ 2J\sqrt{\gamma}q^{[2]}},
\label{g2.10}
\eeq
where $S$ may be also rewritten in terms of each free field by
 the relations
\beqa
S(u)&=&
\left \{ e^{-4\pi p_0^{[1]} \sqrt{\gamma} }
\int_0^uV^{(-1)}_{1}(\rho)d\rho +
\int_u^{2\pi}V^{(-1)}_{1}(\rho)d\rho \right \}
{1\over  e^{-4\pi p_0^{[1]} \sqrt{\gamma} } -1} \quad
\label{g2.11} \\
{-1\over S(u)}&=&
\left \{ e^{-4\pi p_0^{[2]} \sqrt{\gamma} }
\int_0^uV^{(-1)}_{-1}(\rho)d\rho +
\int_u^{2\pi}V^{(-1)}_{-1}(\rho)d\rho \right \}
{1\over  e^{-4\pi p_0^{[2]} \sqrt{\gamma} } -1 }  \quad
\label{g2.12}
\eeqa
Then we may rewrite Eq.\ref{a4.9} as
\beq
V_m^{(J)}= \sqrt{\scriptstyle{2J \choose J+m}} V_{-J}^{(J)}
S^{J+m} =\sqrt{\scriptstyle{2J \choose J+m}} V_{J}^{(J)} (S^{-1})^{J-m}
\label{g2.13}
\eeq
This satisfies the periodicity condition, since
\beq
S(u+2i\pi)=e^{-4\pi p_0^{[1]}\sqrt{\gamma}} S(u).
\label{g2.9}
\eeq
\subsection{Quantization}
One simply replaces\cite{GN3}  the above Poisson brackets by commutators.
Remarkably the above structure carries over to the quantum case.
One may show\cite{GN3}, that for generic $\gamma$,
there exist two equivalent free quantum fields:
\beq
q^{[j]}(x)=
q^{[j]}_0+p^{[j]}_0 x+i\sum_{n\not= 0} e^{-inx}  p^{[j]}_n/n,
\quad p^{[j]}(x)=q^{[j]'}(x).
\label{a4.20}
\eeq
such that
$$
\Bigl [p^{[1]}(\sigma_1,\tau),p^{[1]}(\sigma_2,\tau) \Bigr ]=
\Bigl [p^{[2]}(\sigma_1,\tau),p^{[2]}(\sigma_2,\tau) \Bigr ]
=2\pi i\,  \delta'(\sigma_1-\sigma_2),
$$
\beq
p^{[1]}_0=-p^{[2]}_0,
\label{a4.21}
\eeq
\beq
N^{[1]}\bigl (p^{[1]}\bigr )^2+ p^{[1]'}_1/\sqrt \gamma=
N^{[2]}\bigl (p^{[2]}\bigr )^2+ p^{[2]'}_1/\sqrt \gamma
\label{a4.22}
\eeq
$N^{[1]}$ (resp. $N^{[2]}$)  denote  normal orderings
 with respect to the modes of $p^{[1]}$  (resp. of
$p^{[2]}$).
 Eq.\ref{a4.22} defines the
stress-energy tensor  and the  coupling constant
$\gamma$ of  the quantum theory. The former  generates a
representation of the Virasoro algebra
with central charge $C=3+1/\gamma$.
At an intuitive
level, the
correspondence between $p^{[1]}$ and $p^{[2]}$ may be
understood
from the fact that the Verma modules, which they  generate,
coincide
since   the highest weights only depend upon
$(p_0^{[1]})^2=(p_0^{[2]})^2$.
Next, the quantum version of the differential equation Eq.\ref{2.5}
is derived as follows. Consider the operator
$$
  N^{[j]}\bigl( e^{\alpha q^{[j]}/2} \bigr)
\equiv
e^{\alpha q_0^{[j]}/2}\,e^{ \alpha
p_0^{[j]}x/2}
e^{-ix  \alpha^2/4 }\times
$$
\beq
 \exp\left ((\alpha/2)  i\sum_{n<0}e^{-inx }p_n^{[j]}/ n\right )
\exp\left ((\alpha/2)i\sum_{n>0}e^{-inx}p_n^{[j]}/ n\right ),
\label{a4.23}
\eeq
where $\alpha$ is a constant to be determined. Clearly one has
\beq
{d^2\over dx^2} N^{[j]}\left( e^{\alpha q^{[j]}/2} \right)=
N^{[j]}\left( e^{\alpha q^{[j]}/2} ({\alpha^2 \over 4}
p^{[j]\, 2} +{\alpha\over 2} p^{[j]'} ) \right).
\label{a4.24}
\eeq
On the other hand, the quantum version of  Eqs.\ref{a4.10}  for
the Virasoro
generators are given by
$$
L_n={1\over 2} \left (\sum_r p_r^{[1]} p_{n-r}^{[1]} -in {p_n^{[1]}\over
\sqrt{\gamma}}\right ) =
\left (\sum_r p_r^{[2]} p_{n-r}^{[2]} -in {p_n^{[2]}\over
\sqrt{\gamma}}\right ), \quad n\not = 0
$$
\beq
L_0={1\over 2} \sum_r p_r^{[1]} p_{-r}^{[1]} + {1\over 8 \gamma}
= {1\over 2} \sum_r p_r^{[2]} p_{-r}^{[2]} + {1\over 8 \gamma}.
\label{a4.25}
\eeq
It is straightforward to verify that if $\alpha$ satisfies the equation
\beq
\alpha+{2 \over \alpha}={1\over \sqrt{\gamma}},
\label{a4.26}
\eeq
then Eq.\ref{a4.24} may be rewritten as
$$
{d^2\over dx^2} N^{[j]}\bigl( e^{\alpha q^{[j]}/2} \bigr)+
{\alpha^2\over 2} \Bigl(\sum_{n<0}\,L_n\,e^{-inx}
+{L_0\over 2}+({\alpha^2 \over 32\pi}-{1\over
8 \gamma })\Bigr)N^{[j]}\bigl( e^{\alpha q^{[j]}/2} \bigr)+
$$
\beq
+{\alpha^2\over
2}N^{[j]}\bigl( e^{\alpha q^{[j]}/2} \bigr)
\Bigl(\sum_{n>0}\,L_n\,e^{-inx}
+{L_0\over 2}\Bigr)=0.
\label{a4.27}
\eeq
Since the operators $L_n$ may be equivalently expressed in terms
of $p^{[1]}$ or $p^{[2]}$, it follows that $N^{[j]}\bigl( \exp (\alpha
q^{[j]}/2) \bigr)$ are solution of the same operator differential
equation. Note that, for a given value of $\gamma$, Eq.\ref{a4.26}
which is quadratic gives two values of $\alpha$ we shall denote them
by $\alpha_\pm$. They are identical to the so-called
screening charges of the Coulomb-gas picture.
It is convenient to introduce
two parametres noted $h$, and $\hhat$ such that
\beq
h=\pi \alpha_-^2/2,\quad \hhat =\pi \alpha_+^2/2.
\label{a4.28}
\eeq
They are the quantum-group deformation parameters as we will soon
see. Finally, the quantification of the Schr\"odinger equation gives
us four fields noted $V_j$, and $\Vhat_j$:
\beq
V_j =\, N^{(j)}\bigl (e^{\sqrt{h /2\pi}\>
q^{[j]}}\bigr
), \quad
\Vhat_j =
N^{(j)}\bigl (e^{\sqrt{\hhat /2\pi}\> q^{[j]}}\bigr ),
\quad
j=1,\>2,
\label{a4.29}
\eeq
which obey the differential equations
$$
 -{d^2V_j(x)\over dx^2}+({h\over
\pi})\Bigl(\sum_{n<0}\,L_n\,e^{-inx}
+{L_0\over 2}+({h\over 16\pi}-{C-1\over
24})\Bigr)V_j(x)
$$
\beq
+({h\over
\pi})V_j(x)\Bigl(\sum_{n>0}\,L_n\,e^{-inx}
+{L_0\over 2}\Bigr)=0
\label{a4.30}
\eeq
$$ -{d^2\Vhat_j(x)\over dx^2}+({\hhat\over
\pi})
\Bigl(\sum_{n<0}\,L_n\,e^{-inx}
+{L_0\over 2}+({h\over 16\pi}-{C-1\over
24})\Bigr)\Vhat_j(x)
$$
\beq
+({\hhat\over
\pi})\Vhat_j(x)\Bigl(\sum_{n>0}\,L_n\,e^{-inx}
+{L_0\over 2}\Bigr)=0.
\label{a4.31}
\eeq
These are  operator-Schr\"odinger equations which are the
quantum version of Eq.\ref{2.5}. They are  equivalent
to the decoupling of Virasoro
null-vectors\cite{GN4}.
There are two possible quantum modifications $h$
and
$\hhat$, and thus
there are four solutions. Since $C=1+3/\gamma$, $h$, and $\hhat$
are given by
\beq
h={\pi \over 12}\Bigl(C-13 -
\sqrt {(C-25)(C-1)}\Bigr),\quad
\hhat={\pi \over 12}\Bigl(C-13
+\sqrt {(C-25)(C-1)}\Bigr).
\label{a4.32}
\eeq
By operator product $V_j$, $j=1$, $2$, and $\Vhat
_j$,
$j=1$, $2$,
generate two infinite families of chiral
fields  $V_m^{(J)}$, $-J\leq m \leq J$,
and
$\Vhat_\mhat^{(\Jhat)}$, $-\Jhat\leq \mhat \leq \Jhat$;
  with $V_{-1/2}^{(1/2)}= V_1$,
$V_{1/2}^{(1/2)}= V_2$, and
$\Vhat_{-1/2}^{(1/2)}=
\Vhat_1$ ,
$\Vhat_{1/2}^{(1/2)}= \Vhat_2$.
The fields
$V_m^{(J)}$, $\Vhat_\mhat^{(\Jhat)}$,  are
 of the type ($1$, $2J+1$) and ( $2\Jhat+1$,$1$),
respectively,
in the BPZ
classification. For the zero-modes, it is
simpler\cite{G1}
to  define  the rescaled variables
\beq
\varpi=i p_0^{[1]} \sqrt{{2\pi\over h }}; \quad
\varpihat=i p_0^{[1]} \sqrt{{2\pi \over \hhat }};
\qquad \varpihat=\varpi\>{h\over \pi};
\quad \varpi=\varpihat \>{\hhat\over \pi}.
\label{aA.10}
\eeq
The Hilbert space in which the operators $\psi$ and
$\psihat$
live,
 is a direct sum\cite{G1,G2,G3} of Verma modules
${\cal H}(\varpi)$.
They  are eigenstates of the quasi momentum $\varpi$, and
satisfy $L_n\vert \varpi>= 0$, $n>0$;
$(L_0~-~\Delta(\varpi))\vert \varpi>~=~0$.
The corresponding highest weights  $\Delta (\varpi)$
 may be rewritten as
\beq
\Delta(\varpi)\equiv {1\over
8\gamma}+{(p_0^{[1]})^2\over 2}
={h\over 4\pi}(1+{\pi\over h})^2-{h\over 4\pi}\varpi^2.
\label{aA.11}
\eeq
The commutation relations Eq.\ref{a4.21} are to be supplemented
by the
zero-mode
ones:
\beq
 \bigl[ q^{[1]}_0,\, p^{[1]}_0\bigr]=
\bigl[ q^{[2]}_0,\, p^{[2]}_0\bigr]=i.
\label{azero}
\eeq
It thus follows (see in particular Eq.\ref{a4.23}), that the
fields $V$ and $\Vhat$  shift the quasi momentum
$p^{[1]}_0=-p^{[2]}_0$ by a fixed amount.
For an arbitrary c-number  function  $f$,  one has
\beq
V_m^{(J)}\>f(\varpi)=f(\varpi+2m)\>V_m^{(J)},\quad
 \Vhat_\mhat^{(\Jhat)}\>f(\varpi)=
f(\varpi+2\mhat\, \pi/h)\>\Vhat_\mhat^{(\Jhat)}.
\label{a4.35}
\eeq
The fields $V$ and $\Vhat$ together with
their products noted $V_{m\mhat}^{(J \Jhat)}$
 may thus be naturally restricted to spaces with discrete
values
of $\varpi$.
\section{The OPA of the $U_q(sl(2))$ family.}
In this section, as well as in the next one we only consider
for simplicity only the case of the operators $V_m^{(J)}\equiv
V_{m 0}^{(J 0)}$, which involve  a single screening charge.

\subsection{The fusing and braiding matrices of  the $V$ fields}
The complete operator algebra of the $V$ fields was spelled out
recently\cite{CGR1}  and put in correspondence with the general scheme of
Moore and Seiberg\cite{MS}. We shall  first  describe the result
and give a summary of the derivation later on. The Moore Seiberg (MS)
chiral vertex-operators connect three specified Verma modules
and are thus  of the form
$\phi_{J_3,J_1}^{J_2}$. According to Eq.\ref{a4.35}, the  $V_m^{(J)}$
operators,
on the contrary,
naturally act  in Hilbert
spaces
  of the form
\beq
{\cal H}\equiv
\bigoplus_{n=-\infty}^{+\infty}
{\cal H}(\varpi_0+n).
\label{aA.13}
\eeq
where ${\cal H}(\varpi_0+n)$
are Verma modules.
 $\varpi_0$ is a constant. We shall choose  $\varpi_0=1+\pi/h$, which
corresponds to the
$sl(2,C)$--invariant
vacuum. With this choice, we use the notation ${\cal H}_{J}$
instead of ${\cal H}(\varpi_0+2J)$.
 It is now easy to see that,  according to
Eq.\ref{a4.35},  the  $V$ fields and the MS fields are
related by the projection operator ${\cal P}_{J}$:
\beq
{\cal P}_{J} {\cal H} ={\cal H}_J, \quad
{\cal P}_{J_3} V_m^{(J_2)} \equiv \phi_{J_3,J_3+m}^{J_2}
\label{ar2.16}
\eeq
 The $V$ fields are such that $<\varpi_2 | V_m^{(J)} |\varpi_1>$
is
equal to
one if $\varpi_1=\varpi_3+2m$, and is equal to zero
otherwize. This normalization is  required by the symmetry between
three
legs (sphere with three punctures). From now on,
we restrict ourselves to the
case of genus zero, so that we perform a conformal transformation on the
operator to change from the coordinate $x$ to the coordinate $z=\exp(ix)$.
The   complete
fusion and braiding algebras
take the form:
$$
{\cal P}_{K}\> V^{(J_1)}_{m_1}(z_1)\,V^{(J_2)}_{m_2}(z_2) =
\sum _{J= \vert J_1 - J_2 \vert} ^{J_1+J_2}
 F_{K+m_1,J}\!\!\left[
^{J_1}_{K}
 \> ^{\quad J_2}_{K+m_1+m_2 }
\right]
$$
\beq
\sum _{\{\nu\}}
{\cal P}_{K}\> V ^{(J,\{\nu\})} _{m_1+m_2}(z_2)
<\!\varpi_{J},{\{\nu\}} \vert V ^{(J_1)}_{J_2-J}(z_1-z_2)
\vert \varpi_{J_2} \! >;
\label{ar2.17}
\eeq
$$
{\cal P}_{K}\> V^{(J_1)}_{m_1}(z_1)\,V^{(J_2)}_{m_2}(z_2) =
$$
\beq
\sum _{n_2}
 B_{K+m_1,K+n_2 }\!\!\left[
^{J_1}_{K}
 \> ^{\quad J_2}_{K+m_1+m_2 }
\right]
{\cal P}_{K}\> V ^{(J_2,)} _{n_2}(z_2)
 V ^{(J_1)}_{m_1+m_2-n_2}(z_1)
\label{ar2.18}
\eeq
where $F$ and $B$ (the fusing amd braiding matrices)  have numerical
entries. The notation $|\varpi_{J},{\{\nu\}}>$ represents an arbitrary
state in ${\cal H}_J$.
Using Eq.\ref{ar2.16},
one
may verify
that these expressions
have the general MS form.   In ref.\cite{CGR1}, it was shown that
\beq
F_{{J_{23}},{J_{12}}}\!\!\left[^{J_1}_{J_{123}}
\,^{J_2}_{J_3}\right]
=
{g_{J_1J_2}^{J_{12}}\
g_{J_{12}J_3}^{J_{123}}
\over
g _{J_2J_3}^{J_{23}}\
g_{{J_1}J_{23}}^{J_{123}}
}
\left\{ ^{J_1}_{J_3}\,^{J_2}_{J_{123}}
\right. \left |^{J_{12}}_{J_{23}}\right\}_q.
\label{ar2.19}
\eeq
The  symbol
$\left\{ ^{J_1}_{J_3}\,^{J_2}_{J}
\right. \left |^{J_{12}}_{J_{23}}\right\}_q$  represents the
quantum 6-j
coefficient wich is not completely
tetra\-hed\-ron-symmetric,
as the notation indicates. This is
in contrast with the Racah-Wigner q-6-j symbol, noted
$\left\{ ^{J_1}_{J_3}\,^{J_2}_{J}
\,^{J_{12}}_{J_{23}}\right\}_q$. Recall that the q-6-j symbols are
quantum-group recoupling coefficients which satisfy
the defining relation
$$
\sum_{J_{23}}
\bigl(J_2,M_2;J_3,M_3 \vert J_{23}\bigr)_q
\bigl(J_1,M_1;J_{23},M_{23} \vert J_{123}\bigr)_q
\left \{^{J_1}_{J_3}\,^{J_2}_{J_{123}}\right.
\! \left |^{J_{12}}_{J_{23}}
\right \}_q=
$$
\beq
\bigl(J_1,M_1;J_2,M_2 \vert J_{12}\bigr)_q
\bigl(J_{12},M_{12};J_{3},M_{3} \vert J_{123}\bigr)_q.
\label{2.62}
\eeq
The symbols $\bigl(J_k,M_k;J_\ell,M_\ell \vert J_{k\ell}\bigr)_q$
denote the  q Clebsch-Gordan (3-j)  coefficients\footnote{We use a
condensed notation
$(J_1,M_1;J_2,M_2|J_{12})_q$ instead of
$(J_1,M_1;J_2,M_2|J_{12}, M_1+M_2)_q$.}.

 This term was of course
expected,
in view of the quantum-group structure previously exhibited,
in particular,  in refs.\cite{G1,G3}.
However, there appear, in addition,  coupling constants
$g_{J_1J_2}^{J_{12}}$, which are not trigonometrical functions of
$h$.
Their general expression is
$$
g_{J_1J_2}^{J_{12}}
= \left ({h\over \pi}\right )^{J_1+J_2-J_{12}}
\prod_{k=1}^{J_1+J_2-J_{12}}
\sqrt{
F(1+(2J_1-k+1)h/\pi)\over F(1+kh/\pi)}
$$
\beq
\sqrt{ F(1+(2J_2-k+1)h/\pi)
F(-1-(2J_{12}+k+1)h/\pi)}.
\label{2.39}
\eeq
where $F(z)=\Gamma(z)/\Gamma(1-z)$.
Note that $\Gamma(z)$ is the standard --- not q-deformed ---
Gamma function.
 The result is symmetric in $J_1$, $J_2$. The lack of symmetry
between $J_1$ or $J_2$ and $J_{12}$ is due to the particular
metric
 in the space of primary fields.
The F and B matrices are found to be connected by the MS
relation\cite{MS}
\beq
B^{\pm}_{{J_{23}},{J_{12}}}\!\!\left[^{J_1}_{J}
\,^{J_3}_{J_2}\right]
=e^{\pm i \pi \left( \Delta_{J}+\Delta_{J_2}-\Delta_{J_{23}}
-\Delta_{J_{12}}\right )}
F_{{J_{23}},{J_{12}}}\!\!\left[^{J_1}_{J}
\,^{J_2}_{J_3}\right].
\label{1.2}
\eeq
where $\Delta_J$ is the conformal weight
\beq
\Delta_J=-{h\over \pi} J(J+1)-J
\label{weight}
\eeq
The symbol $\pm$ is chosen according to the ordering of the operator:
taking for instance, $z_i=\exp(i\sigma_1)$, $0\leq \sigma_i \leq \pi$, one
has $\pm=-$ sign $(\sigma_1-\sigma_2)$.
The explicit formula  for the  Racah-Wigner q-6-j coefficients,
which have the tetrahedral symmetry,
are given in
\cite{KR} by
$$
\left\{ ^{a\ b\ e}_{d\ c\ f}\right\}_q=
\big (
\lfloor 2e+1 \rfloor
\lfloor 2f+1 \rfloor
\big )^{-1/2}
(-1)^{a+b-c-d-2e}
\left\{ ^{a\ b}_{d\ c}\right .\left | ^e_f\right\}_q=
$$
$$
=\Delta(a,b,e)\Delta(a,c,f)\Delta(c,e,d)
\Delta(d,b,f)\times
$$
$$
\sum_{z\hbox{ integer}}
(-1)^z
\lfloor z+1 \rfloor \! !
\bigg[
\lfloor z-a-b-e \rfloor \! !
\lfloor z-a-c-f \rfloor \! !
\lfloor z-b-d-f \rfloor \! !\times
$$
\beq
\lfloor z-d-c-e \rfloor \! !
\lfloor a+b+c+d-z \rfloor \! !
\lfloor a+d+e+f-z \rfloor \! !
\lfloor b+c+e+f-z \rfloor \! !
\bigg]^{-1}
\label{2.54}
\eeq
with
$$
\Delta(l,j,k)=
\sqrt{\lfloor -l+j+k \rfloor \! !
\lfloor l-j+k \rfloor \! !
\lfloor l+j-k \rfloor \! !
\over
\lfloor l+j+k+1 \rfloor \! !}
$$
and
\beq
\lfloor n \rfloor \! ! \equiv
\prod_{r=1}^n \lfloor r \rfloor
\qquad \lfloor r \rfloor \equiv {\sin (hr)\over \sin h}.
\label{2.55}
\eeq
In the operator formalism of the Liouville theory, the fusion may be
rewritten as\footnote{We omit the world-sheet variables from now on.}
$$
V^{(J_1)}_{m_1} V^{(J_2)}_{m_2} =
\sum _{J_{12}= \vert J_1 - J_2 \vert} ^{J_1+J_2}
{g_{J_1J_2}^{J_{12}}\>
g_{J_{12}\, ( \varpi-\varpi_0+2m_1+2m_2)/2}
^{\> (\varpi-\varpi_0)/2}
\over
g_{J_{2}\, ( \varpi-\varpi_0+2m_1+2m_2)/2}
^{ \> (\varpi-\varpi_0+2m_1)/2} \>
g_{J_{1}\, ( \varpi-\varpi_0+2m_1)/2}
^{ \> (\varpi-\varpi_0)/2} } \times
$$
$$
\left\{ ^{\quad \quad  J_1}_{( \varpi-\varpi_0 +2m_1+2m_2)/2}
\> ^{\quad J_2}_{( \varpi-\varpi_0)/2}
\right.
\left |^{\quad J_{12}}_{( \varpi-\varpi_0+2m_1)/2}\right\}_q
\times
$$
\beq
\sum _{\{\nu_{12}\}}
V ^{(J_{12},\{\nu_{12}\})}_{m_1+m_2}
<\!\varpi_{J_{12}},{\{\nu_{12}\}} \vert
V ^{(J_1)}_{J_2-J_{12}} \vert \varpi_{J_2} \! >.
\label{2.52}
\eeq
In this last formula, $\varpi$ is an operator. It is easy to
check that this operator-expression
 is equivalent to Eq.\ref{ar2.17}, by computing
the matrix element between the states  $<~\varpi_{J_{123} },
\{\nu_{123}\}|$, and $|\varpi_{J_{3 }}, \{\nu_3\} \! >$. Then,
the additional spins of Eq.\ref{ar2.17}, as compared with
Eq.\ref{2.52}, are given by
\begin{eqnarray}
J_{123}&=&(\varpi-\varpi_0)/2,  \nnn
J_{23}&=&(\varpi-\varpi_0+2m_1)/2,  \nnn
J_{3}&=&(\varpi-\varpi_0+2m_1+2m_2)/2.
\label{2.53}
\eeqa
Similarly, the braiding may be written as
$$
V^{(J_1)}_{m_1} \>  V^{(J_2)}_{m_2} =
\sum_{n_1+n_2=m_1+m_2}
e^{\mp ih (2m_1m_2+m_2^2-n_2^2+\varpi(m_2-n_2))}
\times
$$
$$
\left\{ ^{J_1\quad }_{J_2\quad }
\> ^{( \varpi-\varpi_0 +2m_1+2m_2)/2}
_{ \quad \quad (\varpi-\varpi_0)/ 2}
\right.
\left |^{\quad ( \varpi-\varpi_0 +2n_2)/2}
_{\quad ( \varpi-\varpi_0+2m_1)/2}\right\} \times
$$
\beq
{ g_{J_{1}\, ( \varpi-\varpi_0+2n_1+2n_2)/2}
^{ \> (\varpi-\varpi_0+2n_2)/2} \>
g_{J_{2}\, ( \varpi-\varpi_0+2n_2)/2}
^{ \> (\varpi-\varpi_0)/2}
\over
g_{J_{2}\, ( \varpi-\varpi_0+2m_1+2m_2)/2}
^{ \> (\varpi-\varpi_0+2m_1)/2} \>
g_{J_{1}\, ( \varpi-\varpi_0+2m_1)/2}
^{ \> (\varpi-\varpi_0)/2}} \quad
V^{(J_2)}_{n_2}\>  V^{(J_1)}_{n_1}
\label{2.57}.
\eeq
The correspondence table is again given by Eq.\ref{2.53} with,in addition,
\beq
J_{13}= (\varpi-\varpi_0 +2n_2)/2.
\label{2.58}
\eeq
In the operator-forms Eqs.\ref{2.52}, \ref{2.57}, one sees that
the fusion and braiding matrices involve the operator
$\varpi$, and thus do not commute with the V-operators
(see Eq.\ref{a4.35}). Such is the
general situation of the operator-algebras in the MS formalism.
This is in contrast with, for instance,  the braiding relations
for quantum group representations. In the  article\cite{CGR2}, and
completing the results of refs.\cite{B,G1}, it was shown how to
change
basis to the  holomorphic operators $\xi$ which are
 such that these  $\varpi$
dependences  of the fusing and braiding matrices disappear.
After the transformation,  the fusing and braiding matrices become
equal to the q-Clebsch-Gordan coefficients and to
the universal $R$ matrix, respectively;
and the quantum group  structure becomes  more transparent.
We shall not elaborate on this point here.
\subsection{The polynomial equations}
They basically express the associativity of the operator product
algebra (OPA). First, consider the fusion. The associated pentagonal
equation is derived as follows. We fuse $<\!\varpi_J\vert V^{(J_1)}_{m_1}\,
V^{(J_2)}_{m_2}V^{(J_3)}_{m_3}\vert
\varpi_J+2m_1+2m_2+2m_3\!>$ in two different ways,  beginning
from the left,
and  from the right, and identify the
coefficients of the resulting operator
$V^{(J_{123},\{\nu_{123}\})}_{m_1+m_2+m_3}$.
This gives
$$
\sum_{J_{12}}
F_{J+m_1+m_2,J_{123}}
\!\!\left[
_J^{J_{12}}\,
^{\qquad J_3}_{J+m_1+m_2+m_3}
\right]\,
F_{J+m_1,{J_{12}}}\!\!\left[^{J_1}_J
\,^{\quad J_2}_{J+m_1+m_2}\right]\,\times
$$
$$
\sum_{\{\nu_{12}\}}
<\!\varpi_{J_{123}},\{\nu_{123}\}|
V^{J_{12},\{\nu_{12}\}}_{J_3-J_{123}}|\varpi_{J_3}\!>
<\!\varpi_{J_{12}},\{\nu_{12}\}|
V^{J_1}_{J_2-J_{12}}|\varpi_{J_2}\!>
=
$$
$$
\sum_{J_{23}}
F_{J+m_1,{J_{123}}}\!\left[^{J_1}_J
\,^{\qquad J_{23}}_{J+m_1+m_2+m_3}\right]\,
F_{J+m_1+m_2,{J_{23}}}\!\left[^{\ J_2}_{J
+m_1}\,^{\qquad J_3}_{J+m_1+m_2+m_3}\right]\times
$$
\beq
\sum_{\{\nu_{23}\}}
<\!\varpi_{J_{123}},\{\nu_{123}\}|
V^{J_1}_{J_{23}-J_{123}}|\varpi_{J_{23}},{\{\nu_{23}\}}\!>
<\!\varpi_{J_{23}},\{\nu_{23}\}|
V^{J_2}_{J_3-J_{23}}|\varpi_{J_3}\!>.
\label{2.40}
\eeq
On the r.h.s. we use
\beq
\sum_{\{\nu_{23}\}}\ |\varpi_{J_{23}},{\{\nu_{23}\}}\!>
<\!\varpi_{J_{23}},\{\nu_{23}\}|\
=\ {\cal P}_{{J_{23}}},
\label{2.41}
\eeq
and obtain the factor
$<\!\varpi_{J_{123}},\{\nu_{123}\}|
V^{J_1}_{J_{23}-J_{123}}\
V^{J_2}_{J_3-J_{23}}|\varpi_{J_3}\!>$.
These  last operators are fused in their turn, obtaining
$$
\sum_{J_{23}} F_{J+m_1+m_2,{J_{23}}}\!
\left[^{\ J_2}_{J+m_1}
\,^{\qquad J_3}_{J+m_1+m_2+m_3}\right]
F_{J+m_1,{J_{123}}}\!\left[^{J_1}_J
\,^{\qquad J_{23}}_{J+m_1+m_2+m_3}\right]\,
F_{{J_{23}},{J_{12}}}\!\left[^{J_1}_{J_{123}}
\,^{J_2}_{J_3}\right]
$$
\beq
=\>F_{J+m_1,{J_{12}}}\!\left[^{J_1}_J
\,^{\quad J_2}_{J+m_1+m_2}\right]\,
F_{J+m_1+m_2,{J_{123}}}\!\left[^{J_{12}}_J
\,^{\qquad J_3}_{J+m_1+m_2+m_3}\right].
\label{2.42}
\eeq
On the other hand, the q-6j symbols are well known to satisfy
the same relation\cite{KR}, that is
$$
\sum_{J_{23}}
\left\{ ^{\qquad J_2}_{J+m_1+m_2+m_3}
\,^{\ J_3}_{J+m_1}\right. \left |^{\quad
J_{23}}_{J+m_1+m_2}\right\}_q
\left\{ ^{\qquad J_1}_{J+m_1+m_2+m_3}
\,^{J_{23}}_{J}\right. \left |^{\ J_{123}}_{J+m_1}\right\}_q
\left\{ ^{J_1}_{J_3}\,^{J_2}_{J_{123}}
\right. \left |^{J_{12}}_{J_{23}}\right\}_q
$$
\beq
=
\left\{ ^{\quad J_1}_{J+m_1+m_2}
\,^{J_2}_{J}\right. \left |^{\ J_{12}}_{J+m_1}\right\}_q
\left\{ ^{\qquad J_{12}}_{J+m_1+m_2+m_3}
\,^{J_3}_{J}\right. \left |^{\quad J_{123}}_{J+m_1+m_2}\right\}_q
{}.
\label{2.43}
\eeq
Another example is the Yang Baxter equation
$$
\sum_{J_{134}}
B_{{J_{234}},{J_{134}}}\!\!\left[^{J_{1}}_{J_{1234}}
\,^{J_{2}}_{J_{34}}\right]
B_{{J_{34}},{J_{14}}}\!\!\left[^{J_{1}}_{J_{134}}
\,^{J_{3}}_{J_{4}}\right]
B_{{J_{134}},{J_{124}}}\!\!\left[^{J_{2}}_{J_{1234}}
\,^{J_{3}}_{J_{14}}\right]
$$
\beq
=\sum_{J_{24}}
B_{{J_{34}},{J_{24}}}\!\!\left[^{J_{2}}_{J_{234}}
\,^{J_{3}}_{J_{4}}\right]
B_{{J_{234}},{J_{124}}}\!\!\left[^{J_{1}}_{J_{1234}}
\,^{J_{3}}_{J_{24}}\right]
B_{{J_{24}},{J_{14}}}\!\!\left[^{J_{1}}_{J_{124}}
\,^{J_{2}}_{J_{4}}\right],
\label{2.49}
\eeq
which is obtained from  operator-braidings  in
$<\!\varpi_{J_{1234}}\vert V^{(J_1)}\,V^{(J_2)}\,
V^{(J_3)}\vert \varpi_{J_4}\!>$.
The 6-j symbols also satisfy a similar equation
$$
\sum_{J_{134}}
e^{i\pi\epsilon
(\Delta_{J_{1234}}-\Delta_{J_{124}}
-\Delta_{J_{134}}-\Delta_{J_{234}})}
\left\{ ^{J_{1}}_{J_{2}}\,^{J_{34}}_{J_{1234}}
\right. \left |^{J_{134}}_{J_{234}}\right\}_q
\left\{ ^{J_{1}}_{J_{3}}\,^{J_{4}}_{J_{134}}
\right. \left |^{J_{14}}_{J_{34}}\right\}_q
\left\{ ^{J_{2}}_{J_{3}}\,^{J_{14}}_{J_{1234}}
\right. \left |^{J_{124}}_{J_{134}}\right\}_q
$$
\beq
=\sum_{J_{24}}
e^{i\pi\epsilon
(\Delta_{J_{4}}-\Delta_{J_{14}}-\Delta_{J_{24}}
-\Delta_{J_{34}})}\left\{
^{J_{2}}_{J_{3}}\,^{J_{4}}_{J_{234}}
\right. \left |^{J_{24}}_{J_{34}}\right\}_q
\left\{ ^{J_{1}}_{J_{3}}\,^{J_{24}}_{J_{1234}}
\right. \left |^{J_{124}}_{J_{234}}\right\}_q
\left\{ ^{J_{1}}_{J_{2}}\,^{J_{4}}_{J_{124}}
\right. \left |^{J_{14}}_{J_{24}}\right\}_q,
\label{2.50}
\eeq
Quite generally, the 6-j symbols are solutions of all polynomial
equations. It is quite simple to see that the $gg/gg$ factors are
``pure gauges '' from the viewpoint of MS coditions since they cancel
in pairs. Indeed one may define new chiral vertex operators $\widetilde V$, by
\beq
{\cal P}_{J_{12}} {\widetilde V}_{J_2-J_{12}}^{(J_1)}
\equiv g_{J_1\, J_2}^{J_{12}}
{\cal P}_{J_{12}}  V_{J_2-J_{12}}^{(J_1)},
\label{2.44}
\eeq
so that the fusing and braiding matrices  simply become
$$
{\widetilde F}_{{J_{23}},{J_{12}}}\!\!\left[^{J_1}_{J}
\,^{J_2}_{J_3}\right]
=
\left\{ ^{J_1}_{J_3}\,^{J_2}_{J}
\right. \left |^{J_{12}}_{J_{23}}\right\}_q,
$$
\beq
{\widetilde B}^{\pm}_{{J_{23}},{J_{12}}}\!\!\left[^{J_1}_{J}
\,^{J_3}_{J_2}\right]
=e^{\pm i \pi \left( \Delta_{J}+\Delta_{J_2}-\Delta_{J_{23}}
-\Delta_{J_{12}}\right )}
\left\{ ^{J_1}_{J_3}\,^{J_2}_{J}
\right. \left |^{J_{12}}_{J_{23}}\right\}_q.
\label{2.45}
\eeq
In the associativity conditions for the $\Vt$ operators the $g$'s have
disappeared. Let us stress, however, as already mentioned in the introduction,
that {\bf this does not mean that our fusing and
braiding matrices are equivalent to
Eqs.\ref{2.45}  from the viewpoint of
conformal theory}. Indeed, the new fusing equations  read
$$
{\cal P}_{J}
{\widetilde V}^{(J_1)}_{m_1}\, {\widetilde V}^{(J_2)}_{m_2} =
{\cal P}_{J}
\sum _{J_{12}= \vert J_1 - J_2 \vert} ^{J_1+J_2}
{\widetilde F}_{J+m_1,J_{12}}\!\!\left[
^{J_1}_J
\>^{\quad J_2}_{J+m_1+m_2 }
\right]\times
$$
\beq
\sum _{\{\nu_{12}\}}  {\widetilde V}^{(J_{12},\{\nu_{12}\})} _{m_1+m_2}
<\!\varpi_{J_{12}},{\{\nu_{12}\}} \vert
 {\widetilde V}^{(J_1)}_{J_2-J_{12}} \vert \varpi_{J_2} \! >.
\label{2.46}
\eeq
The matrix element of ${\widetilde V}$ on the right-hand side is a
book-keeping device to recover the coefficients of the OPE.
There  the normalization of ${\widetilde V}$ appears explicitly.
According to Eqs.\ref{2.44},  the matrix elements of $\Vt$ are such that
\beq
<\varpi_{J_2}|{\widetilde V}_m^{(J_1)}(1)|\varpi_{J_{12}}>
= g_{J_1 \, J_2}^{J_{12}}\>
\delta_{m,\,  J_{12} -J_2}.
\label{2.47}
\eeq
For instance, at the level of primaries we have
\beq
{\cal P}_{J}
{\widetilde V}^{(J_1)}_{m_1}\, {\widetilde V}^{(J_2)}_{m_2} =
{\cal P}_{J}
\sum _{J_{12}= \vert J_1 - J_2 \vert} ^{J_1+J_2}
g_{J_1 \, J_2}^{J_{12}}\>
{\widetilde F}_{J+m_1,J_{12}}\!\!\left[
^{J_1}_J
\>^{\quad J_2}_{J+m_1+m_2 }
\right]
 {\widetilde V}^{(J_{12})}
_{m_1+m_2} \> +\cdots .
\label{2.48}
\eeq
The $g$'s have re-appeared. There is no way to get rid of them
at the
level of the two dimensional  OPA.
\subsection{Some ideas about the derivation of the OPA}
It follows from the operator differential-equation Eq.\ref{a4.30} that,
for any holomorphic primary operator
$A_\Delta(x)$
with conformal weight $\Delta$, one has (see Eq.A.6 of
ref\cite{G1}):
$$
< \varpi_4 \vert \, V^{(1/2)}_{\pm 1/2}(x)\, A_\Delta(x')\,
 \vert \varpi_1> =
e^{ix'(\varpi_1^2-\varpi_4^2)h/4\pi}e^{i(x'-x)
(-1/2\mp\varpi_4)h/2\pi }$$
\beq
\times \bigl(1-e^{-i(x-x')}\bigr)^\beta \,
 F(a_\pm ,b_\pm ;c_\pm ;e^{i(x'-x)}),
\label{x2.1}
\eeq
$$a_\pm =\beta -{h\over 2\pi}\mp
h({\varpi_4-\varpi_1 \over 2\pi});\quad
b_\pm =\beta  -{h\over 2\pi}\mp h({\varpi_4+\varpi_1 \over
2\pi});\quad c_\pm =1\mp{h\varpi_4 \over \pi};$$
\beq
\beta ={1\over 2} (1+h/\pi)\bigl(1-
\sqrt{1-{8h\Delta  \over 2\pi(1+h/\pi)^2}}\bigr);
\label{x2.2}
\eeq
where $F(a,b;c;z)$ is the standard hypergeometric function. Moreover
$$
< \varpi_4 \vert \, A_\Delta(x')\,V^{(1/2)}_{\pm 1/2}(x)\,
 \vert \varpi_1> =
e^{ix'(\varpi_1^2-\varpi_4^2)h/4\pi}e ^{i(x-x')
(-1/2\pm \varpi_1)h/2\pi }\times
$$
\beq
\bigl(1-e^{-i(x-x')}\bigr)^\beta \,
 F(a'_\pm ,b'_\pm ;c'_\pm;\,  e^{-i(x-x')}),
\label{y2.3}
\eeq
\beq
a'_\pm =\beta -{h\over 2\pi}\mp h({\varpi_1-\varpi_4 \over
2\pi});\quad
b'_\pm =\beta  -{h\over 2\pi}\mp h({\varpi_1+\varpi_4 \over
2\pi});\quad c'_\pm =1\mp{h\varpi_1 \over \pi}.
\label{x2.4}
\eeq
Consider the fusion. Making use of the well known identity
$$
F(a,b;c;x)= {\Gamma(c)\Gamma(c-a-b) \over \Gamma(c-a)
\Gamma (c-b) } F(a,b;a+b-c+1;1-x)+
$$
\beq
 {\Gamma(c)\Gamma(a+b-c) \over \Gamma(a)
\Gamma (b) }\>
(1-x)^{c-a-b} F(c-a,c-b;c-a-b+1;1-x),
\label{2.14}
\eeq
one derives  Eq.\ref{ar2.17}, with $J_1=1/2$, and
$$
F_{J+\epsilon_1/2,J_2+\epsilon_2/2}\!
\left[^{1/2}_{J}\,^{J_2}_{J_3}\right]=
$$
$$
{\Gamma((1-\epsilon_1)-\epsilon_1(2J+1)h/\pi)
\over
\Gamma(1+(\epsilon_2-\epsilon_1)/2+(J_3+\epsilon_2
J_2-\epsilon_1J+(1-\epsilon_1+\epsilon_2)/2)h/\pi)
}
\times
$$
\beq
{\Gamma(\epsilon_2+\epsilon_2(2J_2+1)h/\pi)
\over
\Gamma((\epsilon_2-\epsilon_1)/2+(-J_3+\epsilon_2
J_2-\epsilon_1J+(-1-\epsilon_1+\epsilon_2)/2)h/\pi)},
\label{2.27}
\eeq
where $\epsilon_i=\pm 1$. Introduce
\beq
A\!\left[ ^{J_1}_{J_3}\,^{ J_2}_{J}
\,^{J_{12}}_{ J_{23}}\right ]\equiv
{F_{{J_{23}},{J_{12}}}\!\!\left[^{J_1}_{J}
\,^{J_2}_{J_3}\right]
\over
\left\{ ^{J_1}_{J_3}\,^{J_2}_{J}
\right. \left |^{J_{12}}_{J_{23}}\right\}_q
}.
\label{2.29}
\eeq
Eq.\ref{2.27} shows that
 $$
A\!\left[ ^{1/2}_{J_3}\,^{ J_2}_J
\,^{J_2+\epsilon_2/2}_{\ J+\epsilon_1/2}\right]
$$
$$
=\sqrt{
F(\epsilon_2+\epsilon_2(2J_2+1)h/\pi)\over
F((1+(\epsilon_2-\epsilon_1)/2)
+(\epsilon_2J_2-\epsilon_1J+J_3+(1-\epsilon_1+\epsilon_2)/2)h/\pi)}
$$
\beq
\times \sqrt{
F((1-\epsilon_1)-\epsilon_1(2J+1))h/\pi)
\over
F((\epsilon_2-\epsilon_1)/2+(\epsilon_2J_2
-\epsilon_1 J-J_3+(-1-\epsilon_1+\epsilon_2)/2)h/\pi)},
\label{2.30}
\eeq
Assume next that the fusing matrix takes the form Eq.\ref{ar2.19}.
Using the explicit expression Eq.\ref{2.54}, this gives
\beq
{
g_{J_2+\epsilon_2/2,\, J_3}^{J}
\over
g _{J_2J_3}^{J+\epsilon_1/2}
}
= {g_{1/2,\, J+\epsilon_1/2}^{J}\over
g_{1/2,\, J_2}^{J_2+\epsilon_2/2}}
A\!\left[ ^{1/2}_{J_3}\,^{ J_2}_{J}
\,^{J_2+\epsilon_2/2}_{ J+\epsilon_1/2}\right].
\label{2.32}
\eeq
Clearly, once
 we know the expression  of $g_{1/2\>J}^{J\pm
1/2}$,
this last relation  will allow us  to determine all of the
$g_{J_1 \, J_2}^{J_{12}}$
by a double recursion on the indices $J_1$ and $J_{12}$ (for any
$J_2$). The basic point is that the integrability condition
for this recurrence relation, completely fixes
$g_{1/2\>J}^{J\pm
1/2}$.
 Note that
$g_{1/2\>J}^{J+1/2}$  can be taken  equal
to one, without loss of generality. For,
if it  were  not, it would be  possible to define
$$
\tilde g_{J_1J_2}^{J_{12}}\equiv
{\alpha_{J_1}\alpha_{J_2}
\over \alpha_{J_{12}}}
g_{J_1J_2}^{J_{12}}
\qquad
\hbox{such\ that}
\qquad
\tilde g_{1/2\ J}^{J+1/2}\equiv
{\alpha_{1/2}\alpha_J
\over \alpha_{J+1/2}}
g_{1/2\ J}^{J+1/2}
=1,
$$
and the fusion coefficients $F$ defined by
Eq.\ref{ar2.19} with $\tilde g$ are the same as the ones
defined with $g$. With this condition, the integrability condition
gives,  up to an irrelevant constant factor (it does not affect $F$):
\beq
g_{1/2\ J}^{J-1/2}
=g_0
\sqrt{F(1+2Jh/\pi)F(-1-(2J+1)h/\pi)}.
\label{2.38}
\eeq
Finally, one  solves  the recurrence relations, and compute
the general expression for $g_{J_1\, J_2}^{J_{12}}$, thereby deriving
Eq.\ref{2.39}.

This completes the derivation of Eq.\ref{ar2.19} for $J_1=1/2$.
The final part of the proof is to use Eqs.\ref{2.42}, and \ref{2.43}
to set up a recurrence on $J_1$, starting from our previous derivation of
the case $J_1=1/2$.  Assume that Eq.\ref{ar2.19} holds for
$J_1\leq K$, and write Eqs.\ref{2.42}, and \ref{2.43}
for $J_1\leq K$ and $J_2\leq K$. In Eq.\ref{2.42},
the last term, that is
$F_{J+m_1+m_2,{J_{123}}}\!\left[^{J_{12}}_J
\,^{\qquad J_3}_{J+m_1+m_2+m_3}\right]$ may have $J_{12}>K$, and
then it is the only one which is not known from our hypothesis.Combining this
equality with the similar one for 6-j symbols
one immediately sees that
$F_{J+m_1+m_2,{J_{123}}}\!\left[^{J_{12}}_J
\,^{\qquad J_3}_{J+m_1+m_2+m_3}\right]$ is also given by
Eq.\ref{ar2.19}. Thus this relation holds for all $J$'s.

\subsection{The general (3D) structure}
In this  section we discuss the general structure of the bootstrap
equations.
In ref.\cite{KR}, quantum-group diagrams were introduced which
involve two different ``worlds'': the ``normal'' one and
the ``shadow'' one.  Adopting  this terminology from now on,
we are going to verify  that the OPA of the $V$  is
in exact correspondence with the shadow diagrams.
At the same time we shall discuss the associated three dimensional
aspect. It corresponds
to the quantum-group version of the Regge-calculus
approach to the discrete
three-dimensional gravity\cite{DisGr}
or to the discussion of ref.\cite{W},  for instance.
In the pictorial representations, we omit the $g$ coefficients.Thus,  we
actually make use of the
operator-algebra  expressed in terms of the $\widetilde V$ fields.
Though of great importance for operator-product expansion,
the coupling constants $g$  define a pure gauge for the  polynomial equations
or knot-theory viewpoints.
We could draw other figures including $g$ coefficients,
to show how they cancel out of those equations,
but this  would be cumbersome. The basic fusing and braiding operations
on the $\widetilde V$ operators  have three equivalent representations
$$
\epsffile{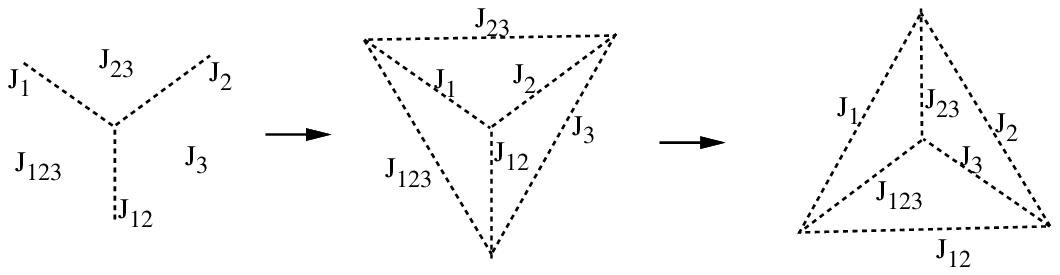}
$$
\beq
=\left\{ ^{J_1}_{J_3}\,^{J_2}_{J_{123}}
\right. \left |^{J_{12}}_{J_{23}}\right\}
\hbox{  (fusion of $\widetilde V$ operators)},
\label{4.1}
\eeq
$$
\epsffile{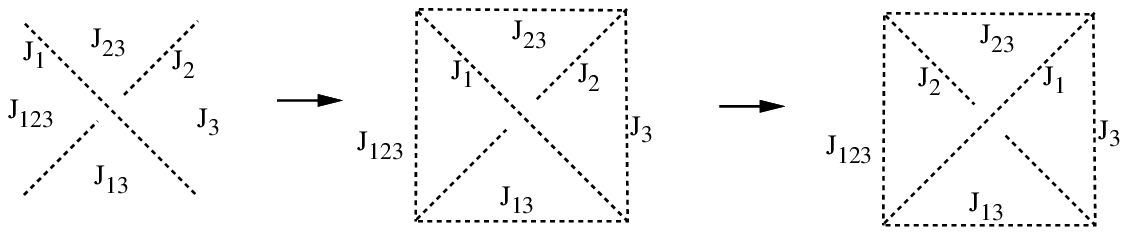}
$$
\beq
= e^{i\pi (\Delta_{J_{123}}+\Delta_{J_3}
-\Delta_{J_{23}}-\Delta_{J_{13}})}
\left\{
^{J_1}_{J_2}\,^{J_3}_{J_{123}}
\right. \left |^{J_{13}}_{J_{23}}\right\}
\hbox{  (braiding of $\widetilde V$ operators)}.
\label{4.2}
\eeq
First consider the left diagrams, and the associated Eqs.\ref{ar2.17},
and \ref{ar2.18}. Apart from the  $\widetilde V$-matrix element
on the right-hand side
of the fusing relation, which has no specific representative, each
operator
$\widetilde V^{(J)}_m$ is represented by a dashed line carrying the
label $J$.
 The spins on the faces display the zero-modes of the
Verma modules on which the  $\widetilde V^{(J)}_m$ operators  act.
Thus the $m$'s are differences between the spin-labels of the two
neighbouring faces. For the braiding  diagram, the spins on the
edges are unchanged at crossings, and, for given $J_1$, $J_2$,
 the braiding
diagram has the form of  a  vertex of
an interaction-around-the-face (IRF) model.
These diagrams  are two-dimensional (2D). The appearence of spins on the faces
reflect the fact that
the  fusion and braiding properties
depend upon the Verma module on which the operators  act.
It is easily seen that, when they are used as building blocks,
  the above
 drawings
generate diagrams which
  have the same structure as the quantum-group  ones of  ref.\cite{KR}
  in the shadow world\footnote{we used
 dashed lines to agree with the conventions of ref.\cite{KR}.}.
The polynomial equations can be viewed as
link-invariance conditions.
For instance,
\begin{equation}
\epsffile{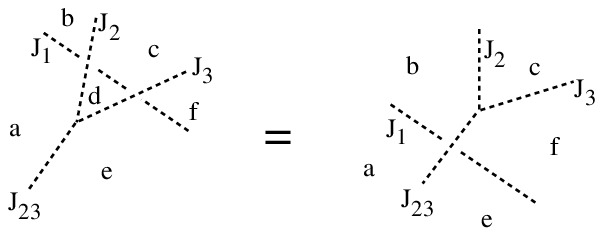}
\label{4.3}
\end{equation}
gives the pentagonal relation
of the $V$ fields discussed in ref.\cite{CGR1},
 after cancellation of the phases
(with a change of indices).

The middle diagrams of figures \ref{4.1}, and \ref{4.2}
 are obtained from the left ones  (first arrow)
by  enclosing the 2D  figures with  extra dashed
lines  carrying  the spin labels
which were previously on the faces. In this way, one gets three-dimensional
(3D)
tetrahedra, with spin labels only on the edges.
The right figures are obtained from the middle ones
by  dualisation:
the  face, surrounded by the  edges $J_a,J_b,J_c$,
becomes the  vertex where  the  edges $J_a,J_b,J_c$ join,
and conversely, a vertex becomes a face.
An edge joining two vertices becomes
the  edge between the two dual  faces.
There is one triangular face
for each $ \widetilde V$ field, including the $\widetilde V$
matrix-element of the
fusing relation Eq.\ref{ar2.17}.
On the dualised polyhedra,
the triangular inequalities give the addition rules for spins.
The main point of the middle and right diagrams is that, as a consequence
of  the basic MS
properties
of the OPA,     they are really 2D projections of
 three-dimensional diagrams which may be rotated at
essentially no cost\footnote{We use
6-j symbols which do not have the full tetrahedral symmetry,
so that two edges should be distinguished.
This is  discussed in ref.\cite{CGR1}.}.
For instance,
 the MS
relation  between fusing and braiding matrices
simply corresponds  to the fact that they are
represented by tetrahedra which may be identified  after  a
rigid 3D
rotation.  We shall illustrate the general properties of the
3D diagrams on the example of the pentagonal relation.
In the same way as
 we closed the basic figures in Eq.\ref{4.1}, \ref{4.2},
the rule to go to 3D  is to close the composite figure \ref{4.3}.
It gives a polyhedron
 which  has vertices with three edges only, which we call type
 V3E.
The two-dimensional Eq.\ref{4.3} now simply corresponds to
viewing the V3E  polyhedron  from two different angles:
\begin{equation}
\epsffile{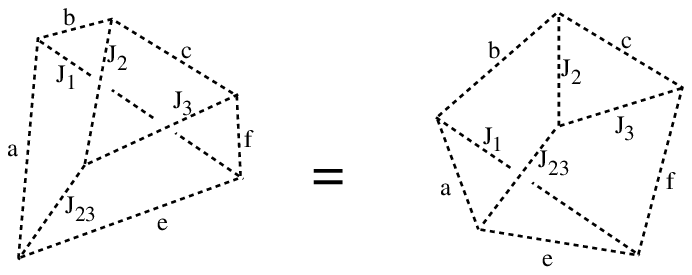}
\end{equation}
Dualisation gives a polyhedron,
with  only triangular faces, which we call F3E.
The polynomial equations are recovered
by decomposing a F3E polyhedron  into  tetrahedra
(this correspondence only works with  F3E
polyhedra, this is the reason for  dualisation).
In parallel with the  two different fusing-braiding
decompositions of each side of
Eq.\ref{4.3},
there are  two  3D  decompositions of the F3E  polyhedron.
This is represented in split view on the next figure,
where the internal
faces are  hatched for clarity.
\begin{equation}
\epsffile{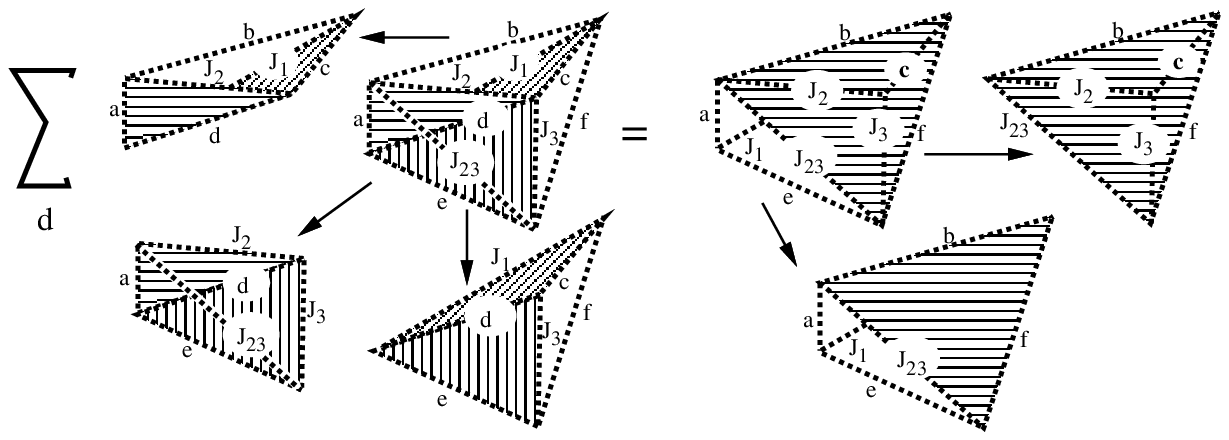}
\label{4.5}
\end{equation}
In general,  the rule is to take a polyhedron with triangular faces
and to decompose it in tetrahedra in different ways.
Substituting the associated 6-j symbols  yields the polynomial
identities\footnote{we restrict ourselves to
polyhedra which are orientable surfaces.}.
In quantum-group diagrams\cite{KR}, a second world was introduced -- the
normal one -- which is represented by solid lines.
 It corresponds to the $\xi$ fields which we leave out of the present
discussion.
\section{Continuation to continuous $J$}
The above discussion should be extended to non integer $2J$ in order
to achieve a complete description of 2D gravity. This appears in
several  instances. First, the  Liouville
exponentials
$e^{-J\alpha_- \Phi}$  with half integer $J$  can be obtained
from the discussion  just recalled, but this does not allows to
reach the field $\Phi$ itself. It may only be defined as
$\Phi =-{d\over dJ}|_{J
=0} \exp (-J\alpha_- \Phi)/ \alpha_- $, if continuous $J$ may be
handled. Second, in the strong coupling regime, values of
$J$ appear\cite{G3,GR1}  that are fourths or sixths of integers, in
some cases. Treating these  rational
values is tantamount to going to the continuous case.
The case of integer $2J$ corresponds, in the BPZ framework, to the
appearance of degenerate fields satisfying null vector differential
equations. For continuous $J$ these equations are lacking.
However,  the structural analogy of $
U_q(sl(2))$
with its classical counterpart together with the group-theoretical
decomposition of the classical Liouville exponentials suggests
that
in the general situation, the chiral primaries should fall into one-sided
highest/lowest weight representations of $ U_q(sl(2))$. Furthermore
one expects that they should obey a closed algebra under fusion and
braiding, determined by the q-Clebsch-Gordan coefficients resp.
R-matrix of $ U_q(sl(2)$ relevant for these representations.
By setting up a
Coulomb-gas type representation for the chiral vertex operators with
arbitrary real $J$, their exchange algebra\index{exchange algebra}
 becomes accessible to
free field techniques, and we can prove that the braiding matrix is
given by a natural analytic continuation of the positive half-integer
spin case, defined in terms of Askey-Wilson polynomials. Then, the
fusion matrix is determined by a generalization of the pentagonal
equations discussed earlier.
This part is a summary of refs.\cite{GS1,GS3}.
\subsection{The braiding}
The purpose of this section is to derive that the operators
$V_m^{(J)}$ ($\equiv V_{m, 0}^{(J, 0)}$)
 with $J+m=0,1,2,\dots$ and continuous $J$,
 obey a closed exchange algebra. The classical
expression
Eq.\ref{g2.11}  for $S$ has a rather simple quantum
generalization, which we
will denote by the $S$ to signify that it is
is a primary field of dimension zero (``screening charge''),
namely\cite{LuS}
\beq
 S(\sigma)= e^{2ih(\varpi+1)}
 \int _0 ^{\sigma} d\rho V_{1}^{(-1)}(\rho)+
 \int _\sigma  ^{2\pi } d\rho V_{1}^{(-1)}(\rho)
\label{3.1}
\eeq
Apart from an overall change of normalization ---
removal of the denominator ---  and the introduction of normal orderings,
 the only change
 consists in the replacement
$\varpi \to \varpi+1$ in the prefactor of the first integral.
 The quantum formula is such that $S$ is
periodic up to a multiplicative factor
\beq
S(\sigma+2\pi)=e^{2ih(\varpi+1)}S(\sigma+2\pi).
\label{3.2}
\eeq
This is the quantum version of Eq.\ref{g2.9}.
The basic primary field of the Coulomb gas picture is defined
as
\beq
U_{m}^{(J)}(\sigma)=  V_{-J}^{(J)}(\sigma) [S(\sigma)]^{J+m}
\label{3.3}
\end{equation}
which is the quantum version of the first equality in Eq.\ref{g2.13}.
The conformal dimension of $U_m^{(J)}$  agrees
with Eq.\ref{weight}. Furthermore,
one easily verifies that
\begin{equation}
U_m^{(J)} \varpi = (\varpi+2m) U_m^{(J)}
\label{3.4}
\end{equation}
which is the same as the first relation of
Eqs.\ref{a4.35}.
Since  conformal weight and zero mode shift define
a primary field uniquely up to a $\varpi$-dependent normalization,
one certainly has
\beq
U^{(J)}_m =I^{(J)}_m(\varpi) V^{(J)}_m
\label{3.5}
\eeq
The explicit expression  for $I^{(J)}_m(\varpi)$ is
\[
I_m^{(J)}(\varpi )=
\left( 2 \pi \Gamma(1+{h\over\pi}) \right )^n
e^{( ih(J+m)(\varpi-J+m))}
\]
\beq
\prod_{\ell=1}^{J+m}
{\Gamma[1+(2J-\ell+1)h/\pi] \over \Gamma[1+\ell h/\pi]
 \Gamma[1-(\varpi+2m-\ell)h/\pi]
\Gamma[1+(\varpi+\ell)h/\pi] }.
\label{3.6}
\eeq
 This formula illustrates
an important
point to be made about the integral representation Eq.\ref{3.1}.
For small enough $h$, the arguments of the gamma functions are all positive,
and this corresponds to the domain where the integral representation is
convergent. When $h$ increases, divergences  appear. However,
the equation
just written continues to make sense provided none  of the arguments  of
gamma functions precisely
 vanish or become  a negative integer. This possibilty is
ruled out generically for continuous $J$ and $\varpi$. Thus there exists
a continuation of the integral representation that makes perfect sense,
and Eq.\ref{3.3} is  meaningful for any $h$.
We shall use the fields $\Vt$ instead of the $V$'s.
The expression for the coupling constants $g$ recalled above has an  immediate
extension to continuous $J$ with $J+m$ integer,
that is
$$
g_{J, \, (\varpi-\varpi_0)/2 +m}^{(\varpi-\varpi_0)/2}
= ({h\over \pi})^{J+m} \times
$$
\beq
\prod_{k=1}^{J+m}
\sqrt{
F[1+(2J-k+1)h/\pi]
F[(\varpi+2m-k)h/\pi]
F[-(\varpi+k)h/\pi]
\over
F[1+kh/\pi]
}.
\label{3.7}
\eeq
The treatment of the
square roots require some care. We follow the prescription of
ref.\cite{G5}
also used in ref.\cite{GR1}. Thus  Eq.\ref{2.44} immediately extends to
the case of non integer $J$:
$\Vt^{(J)}_m=g_{J, \, (\varpi-\varpi_0)/2 +m}^{(\varpi-\varpi_0)/2}
V^{(J)}_m$. The relation with $U^{(J)}_m$ is given by
\beq
U^{(J)}_m ={I^{(J)}_m(\varpi)\over
g_{J, \, (\varpi-\varpi_0)/2 +m}^{(\varpi-\varpi_0)/2}} \Vt^{(J)}_m
\equiv {1\over \kappa_{J, \, (\varpi-\varpi_0)/2 +m}^{(\varpi-\varpi_0)/2}}
\Vt^{(J)}_m.
\label{3.8}
\eeq
After some calculation, one finds
$$
\kappa_{J_1 J_2}^{J_{12}}=
\left ({\pi e^{-i(h+\pi)}
\over 2\Gamma(1+h/\pi) \sin h}\right )^{J_1+J_2-J_{12}}
e^{ih (J_1+J_2-J_{12})( J_1-J_2-J_{12})} \times
$$
\beq
\prod_{k=1}^{J_1+J_2-J_{12}}
\sqrt{ \lfloor 1+2J_1-k\rfloor \over
\lfloor k \rfloor \, \lfloor 1+2J_2-k \rfloor\,
\lfloor 1+2J_{12}+k \rfloor },
\label{3.9}
\eeq
which makes  sense for continuous $J$'s provided $J_1+J_2-J_{12}$ is
a non-negative integer. Note that, for the fields $V_{-J}^{(J)}$,
there  is
no distinction between $\Vt_{-J}^{(J)}$ and $V_{-J}^{(J)}$, since
$g_{J, x-J}^J=1$. Hence, in Eqs.\ref{3.1}, and \ref{3.3}, we may use
$\Vt$ fields to represent normal ordered exponentials, so that
only $\Vt$ fields appear in the discussion.

We now come to
 the braiding  algebra of the fields
$U_m^{(J)}$.
We shall recall some basic points of the derivation following
refs.\cite{GS1,GS3}.
The braiding relation takes the form
\beq
U_m^{(J)}(\sigma) U_m^{(J')}(\sigma')=
\sum_{m_1,m_2} R_U(J,J';\varpi)^{m_2 m_1}_{m\phantom{_2} m'\phantom{_1}}
U_{m_2}^{(J')}(\sigma') U_{m_1}^{(J)}(\sigma)
 \label{3.10}.
\eeq
We only deal with the case $\pi >\sigma '>\sigma >0$ explicitly. The other
cases are deduced from the present one in the standard way.
The sums extending over non-negative integer $J+m_1$ resp. $J'+m_2$.
with the condition
\begin{equation}
m_1+m_2=m+m'=:m_{12}.
\label{3.11}
\end{equation}
Since one considers the braiding at
equal $\tau$ one lets $\tau=0$ once and for all, and works  on the unit
circle $u=e^{i\sigma}$.
As there are no  null-vector decoupling equations for continuous $J$,
the derivation of Eq.\ref{3.10}  is only based   on the free
field techniques summarized  in section 2. The basic point of our
derivation
is that
 the exchange of two $U_m^{(J)}$
operators can be mapped into an equivalent problem in one-dimensional
quantum mechanics, and becomes just finite-dimensional linear
algebra.
In  view of Eq.\ref{3.1}, \ref{3.3},
 the essential observation
is  that one only needs  the braiding relations of
$\Vt_{-J}^{(J)}$ operators which are normal ordered exponentials.
(``tachyon operators''). One may verify that
$$
\Vt_{-J}^{(J)}=N^{[1]}\left ( e^{2J \sqrt{h/2\pi} q^{[1]}}\right),
$$
where we used the notation of section 2. This formula
clearly makes sense for arbitrary $J$. An elementary computation
 gives
\begin{equation}
\Vt_{-J}^{(J)}(\sigma )\Vt_{-J'}^{(J')}(\sigma ')=
e^{-i2JJ' h \epsilon (\sigma -\sigma')}
\Vt_{-J'}^{(J')}(\sigma ' )\Vt_{-J}^{(J)}(\sigma )
\label{3.12}
\end{equation}
where $\epsilon (\sigma -\sigma')$ is the sign of $\sigma -\sigma'$.
This means that when commuting the tachyon operators in
$U_{m'}^{(J')}(\sigma ')$ through those of
$U_{m}^{(J)}(\sigma )$, one only encounters phase factors of the form
$e^{\pm i2 \alpha \beta h } $ resp. $e^{\pm  6i \alpha \beta h } $,
with $\alpha$ equal to $J$ or $-1$, $\beta$ equal to $J'$ or $-1$,
since  we take
$\sigma, \sigma' \in  [0,\pi]$. Hence we are led to decompose
the
integrals defining the screening charges S into pieces which commute
with each other and  with $\Vt_{-J}^{(J)}(\sigma)$, $\Vt_{-J'}^{(J')}
(\sigma ')$ up to one of the above phase factors.
Let us write
\[
S(\sigma )\phantom{'} = S_{\sigma \sigma '} + S_\Delta,
\quad
S(\sigma ')= S_{\sigma \sigma '} + k(\varpi)S_\Delta \equiv
S_{\sigma \sigma '} + \tilde {S}_\Delta,
\]
\[
S_{\sigma \sigma '}:= k(\varpi)
\int_0^\sigma \Vt_{1}^{(-1)}(\rho )
d\rho + \int_{\sigma '}^{2\pi}\Vt_{1}^{(-1)}(\rho )
d\rho,
\]
\begin{equation}
S_\Delta := \int_\sigma^{\sigma '}\Vt_{1}^{(-1)}(\rho )
d\rho,
\quad
k(\varpi):= e^{2ih(\varpi+1)}
\label{3.13}
\end{equation}
Using Eq.\ref{3.12}, we then get the following simple algebra
for
$S_{\sigma \sigma '}, S_\Delta , \tilde {S}_\Delta$:
\medskip
\begin{equation}
S_{\sigma \sigma '}S_\Delta =q^{-2}S_\Delta S_{\sigma \sigma '},
\quad
S_{\sigma \sigma '}\tilde {S}_\Delta =q^2 \tilde {S}_\Delta
S_{\sigma \sigma '},
\quad
S_\Delta \tilde {S}_\Delta =q^4 \tilde {S}_\Delta S_\Delta,
\label{3.14}
\end{equation}
and their commutation properties with
$\Vt_{-J}^{(J)}(\sigma ),\Vt_{-J'}^{(J')}(\sigma ' )$ are given by
 \begin{equation}
\begin{array}{lll}
&\Vt_{-J}^{(J)}(\sigma) S_{\sigma \sigma '}=q^{-2J}S_{\sigma \sigma'}
\Vt_{-J}^{(J)}(\sigma ),
&\Vt_{-J'}^{(J')}(\sigma')S_{\sigma \sigma '}= q^{-2J'} S_{\sigma \sigma'}
\Vt_{-J'}^{(J')}(\sigma'),
\nonumber\\
&\Vt_{-J}^{(J)}(\sigma)S_\Delta  = q^{-2J}S_\Delta \Vt_{-J}^{(J)}(\sigma ),
&\Vt_{-J}^{(J)}(\sigma) \tilde {S}_\Delta \  =q^{-6J} \tilde {S}_\Delta
\Vt_{-J}^{(J)}(\sigma),
\nonumber \\
&\Vt_{-J'}^{(J')}(\sigma')S_\Delta = q^{2J'} S_\Delta \Vt_{-J'}^{(J')}
(\sigma '),
&\Vt_{-J'}^{(J')}(\sigma')\tilde {S}_\Delta = q^{-2J'} \tilde
{S}_\Delta
\Vt_{-J'}^{(J')}(\sigma').
\end{array}
\label{3.15}
\end{equation}
Finally, all three screening pieces obviously shift the zero mode in the
same way:
\begin{equation}
 \left. \begin{array}{ccc}
S_{\sigma \sigma '}\nonumber \\ S_\Delta \nonumber \\ \tilde {S}_\Delta
\end{array} \right \} \varpi
= (\varpi +2) \left \{ \begin{array}{ccc}
S_{\sigma \sigma '}\nonumber \\ S_\Delta \nonumber \\ \tilde {S}_\Delta
\end{array} \right. .
\label{3.16}
\end{equation}
Using Eqs.\ref{3.15} we can commute $\Vt_{-J}^{(J)}(\sigma  )$ and
$\Vt_{-J'}^{(J')}(\sigma  ')$ to the left on both sides of  Eq.\ref{3.10},
so that they can be cancelled. Then we are left with
\[
(q^{-2J'}S_\Delta + q^{2J'}S_{\sigma \sigma '} )^{J+m}
(\tilde {S}_\Delta +S_{\sigma \sigma '})^{J'+m'}q^{2JJ'}=
\]
\begin{equation}
\sum_{m_1,m_2} R(J,J';\varpi + 2(J+J'))_{m_{\phantom{2}}m'}^{m_2 m_1}
(q^{2J} S_{\sigma \sigma '} + q^{6J}\tilde {S}_\Delta )^{J'+m_2}
(S_{\sigma \sigma '} + S_\Delta )^{J+m_1}
\label{3.17}
\end{equation}
It is apparent from this equation that the braiding problem of the
$U_m^{(J)}$ operators is governed by the Heisenberg-like algebra
Eq.\ref{3.14}, characteristic of one-dimensional
quantum mechanics. However,
to see this structure emerge, we had to decompose the screening charges
$S(\sigma ), S(\sigma ')$ in a way which depends on both positions
$\sigma ,\sigma '$; hence the embedding of this Heisenberg algebra
into the 1+1 dimensional field theory is somewhat nontrivial.
We shall  proceed using  the following simple representation
of the algebra Eq.\ref{3.14} in terms of one-dimensional
quantum mechanics ($y$ and $y'$ are arbitrary complex numbers):
\begin{equation}
S_{\sigma \sigma '}=y' e^{2Q}, \quad
S_\Delta =y e^{2Q-P}, \quad  \tilde {S}_\Delta =
y e^{2Q+P},
 \qquad [Q,P]=ih  .
\label{3.18}
\end{equation}
The third  relation in Eq.\ref{3.18} follows from
the second one in view of
$\tilde {S}_\Delta =k(\varpi )S_\Delta$ (cf. Eq.\ref{3.13}).
        This means we are
identifying here $P \equiv ih\varpi $ with the zero mode of the original
problem. Using $e^{2Q+cP} = e^{cP} e^{2Q} q^c$ we can commute
all factors
$e^{2Q}$ to the right on both sides of Eq.\ref{3.17} and then
cancel them.
This leaves us with
$$
q^{2JJ'}
\prod_{s=1}^{J+m}(y'q^{2J'} +yq^{-(\varpi -2J+2s-1)})
\prod_{t=1}^{J'+m'} (y'+yq^{\varpi -2J'+2m+2t-1})=
$$
$$
\sum_{m_1} R_U(J,J';\varpi)_{m_{\phantom {2}} m'}^{m_2 m_1}
$$
\beq
\prod_{t=1}^{J'+m_2}(y'q^{2J} +yq^{\varpi +4J-2J'+2t-1})
\prod_{s=1}^{J+m_1} (y'+yq^{-(\varpi -2J+2m_2 +2s -1)})
\label{3.19}
\end{equation}
where we have shifted back $\varpi +2(J+J') \rightarrow \varpi$
compared to Eq.\ref{3.17}.
Since the overall scaling $ y\rightarrow \lambda y,
y' \rightarrow \lambda y'$ only gives back Eq.\ref{3.11}, we can set $y'=1$.

The solution of these equations, which was
 derived in ref.\cite{GS1}, will be
cast under the convenient form
\beq
R(J,J',\varpi)_{m m'}^{m_2 m_1}=
e^{-i\pi (\Delta(c)+\Delta(b)
-\Delta(e)-\Delta(f))}
{\kappa_{ab}^e \kappa_{de}^c\over \kappa_{db}^f \kappa_{af}^c}\left\{
^{a}_{d}\,^{b}_{c}
\right. \left |^{e}_{f}\right\}
\label{3.20}
\eeq
where
$$
a=J,\quad b=x+m+m',\quad c=x\equiv (\varpi-\varpi_0)/2
$$
\beq
d=J',\quad e=x+m_2,\quad f=x+m.
\label{3.21}
\eeq
The r.h.s. may be expressed in terms of q-hypergeometric functions
by the formula
$$
{\kappa_{ab}^e \kappa_{de}^c\over \kappa_{db}^f \kappa_{af}^c}\left\{
^{a}_{d}\,^{b}_{c}
\right. \left |^{e}_{f}\right\}_q
=
$$
\beq
{\lfloor \rho\rfloor_{n'}
\lfloor 1-\beta\rfloor_{n-n_1}
\lfloor \phi \rfloor_{n_1}
 \lfloor 1-\epsilon-n_1\rfloor_{n_1} \over
\lfloor n_1 \rfloor \! !
\lfloor \beta-\epsilon-n'-n \rfloor_{n+n'}
 }
\>_4 F_3\left (^{\alpha,}_{\epsilon,} \,\,
^{\beta ,}_{\phi ,} \,\, ^{-n_1 ,}_
{\rho } \,\, ^{-n'} \ ;q,1 \right )
\label{3.22}
\eeq
where
$$
\alpha=-a-c+f,\quad \beta=-c-d+e,
$$
$$
n_1=a+b-e, \quad n_2=d+e-c,\quad  n=f+a-c \quad n'=b+d-f;
$$
\beq
\epsilon=-(a+b+c+d+1), \quad
\phi=1+n-n_1, \quad
\rho=e+f-a-d+1.
\label{3.23}
\eeq
Explicitly one has
\beq
n_1=J+m_1, \quad
n_2=J+m_2, \quad
n=J+m, \quad
n'=J+m'.
\label{3.24}
\eeq
Since  they are equal to the screening numbers,
they  are positive integers. It follows that Eq.\ref{3.22} makes sense,
since one defines, as in the previous work along the same line,
\[
_4 F_3 \left (^{a,}_{e,} \, ^{b,}_{f,} \,
^{c,}_{g} \, ^{d} ;q,\rho \right )=
\sum_{n=0}^\infty {\lfloor a\rfloor_n \lfloor b\rfloor_n
\lfloor c\rfloor_n \lfloor d\rfloor_n \over \lfloor e\rfloor_n\lfloor
f\rfloor_n \lfloor g\rfloor_n \lfloor n\rfloor !} \ \rho^n ,
\]
\beq
\lfloor a\rfloor_n := \lfloor a\rfloor \lfloor a+1\rfloor \cdots
\lfloor a+n-1\rfloor ,
\quad \lfloor a\rfloor_0 :=1.
\label{3.25}
\eeq

The method used to derive Eq.\ref{3.19} was to transform this
hypergeometric function  into another one such that the desired relations
follow from the orthogonality relation of the associated Ashkey-Wilson
polynomials. In this connection, let us simply note that a
simple reshuffling of the parameters of the latter form,
 allows to verify that
the usual orthogonality relations of the 6-j symbols extends
to our case. One has,  in general,
\beq
\sum_{J_{23}}
\left\{ ^{J_1 }_{J_3 }
\> ^{J_2 } _{ J_{123}}
\right.
\left |^{ J_{12}}
_{J_{23}}\right\}_q
\left\{ ^{J_1 }_{J_3 }
\> ^{J_2 } _{ J_{123}}
\right.
\left |^{ K_{12}}
_{J_{23}}\right\}_q =\delta_{J_{12}-K_{12}}
\label{x3.25}
\eeq
where the $J$'s are arbitrary chosen so that
  the screening numbers
$$
n_1=J_1+J_2-J_{12}, \quad n_2=J_3+J_{12}-J_{123}, \quad
n=J_1+J_{23}-J_{123},
$$
\beq
n'=J_2+J_3-J_{23},\quad
\tilde n_1=J_1+J_2-K_{12}, \quad \tilde n_2=J_3+K_{12}-J_{123},
\label{3.26}
\eeq
are positive integers. These  conditions fix the range of summation over
$J_{23}$.
\subsection{Generalization of  6-j symbols}
In this section, following ref.\cite{GR1},
 we systematize the  generalization of the 6-j symbols
to non-half-integer spins,
and prove the corresponding generalized polynomial equations\footnote{
This is an abusive use of the  name
``polynomial equations'' which usually refers to
consistency equations for  fusion
and braiding coefficients.
But the 6-j coefficients, which  are  solutions  of them,
satisfy parallel equations,
namely orthogonality, Racah identity,
Bidenharn-Elliot identity...,  that we generically
call polynomial equations as well.}
in particular the pentagonal relation. This will allow to complete the
picture of the OPA for continuous $J$, by including the fusion.
For this,  we note that relevent notion is not whether some spins
are half-integer or not, but how many triangular
inequalities are relaxed.
Explicitly, we use dotted vertices, a dot on one leg meaning that
the sum of the spins of the two other legs minus the spin of this leg
is constrained to be a positive integer. Since there are three
legs
at a vertex we may have one two or three dots. We call them
type TI1, TI2, TI3, where  the letters
T and I stands for triangular inequalities. A example of the
TI1 case are
$$
\epsffile{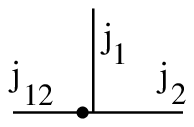}
\ \ \to\ \
j_1+j_2-j_{12}
\hbox{  positive integer}
$$
Adding dots to a vertex adds other restrictions.
For the type TI3, we have
\beq
\epsffile{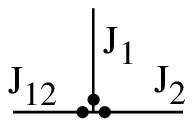}
\to\!
\left\{
\begin{array}{ccc}
J_1+J_2-J_{12} && \!\!\!\hbox{positive integer}\\
J_{12}+J_2-J_1 && \!\!\!\hbox{positive integer}\\
J_1+J_{12}-J_2 && \!\!\!\hbox{positive integer}\\
\end{array}
\right\}
\!\!\Rightarrow
\begin{array}{c}
2J_1,2J_2,2J_{12} \nnn
\hbox{ positive integers}.
\end{array}
\label{ThreeCond}
\eeq
These are but the usual (full) triangular inequalities,
or the branching rules for $U_q(sl(2))$, with half-integer spins.
So, the standard operators  are
of the   TI3 type.

The first step of generalization is to relax one restriction, which give
the TI2 type:
\beq
\epsffile{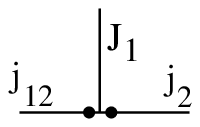}
\ \to\
\left\{
\begin{array}{ccc}
J_1+j_2-j_{12} \hbox{ positive integer}\\
J_1+j_{12}-j_2 \hbox{ positive integer}\\
\end{array}
\right\}
\ \Rightarrow\
2J_1\hbox{ positive integer}
\label{TwoCond}
\eeq
$j_{12}$ and $j_2$ being arbitrary\footnote{For the time being, we denote
half-integer positive spins by capital letters and
continuous  ones by small letters, but this is only a consequence
of the type of vertex and in no case
an a priori assumption.}.
In this TI2 case,  $J_1$ is a positive half-integer, and
$j_2-j_{12}$ is a half-integer (positive or negative).
This is fixed by  the number of restrictions.
The second step is of course to keep only one restriction (type TI1):
\beq
\epsffile{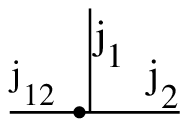}
\ \to\
j_1+j_2-j_{12} \hbox{ positive integer}
\label{OneCond}
\eeq
and none of the spins are half-integers.
Then the fusion or braiding of such generalized vertices lead
to 6-j symbols generalized in a very specific way,
and the ``miracle'' is that it is mathematically consistent.
In the first step of generalization,
fusion and braiding lead to two different
generalized 6-j, whereas they lead to only one kind of 6-j
in the second step.
We define now generalized 6-j coefficients for the fusion
and braiding of operators
with only TI1 conditions. The diagrams are
\beq
\epsffile{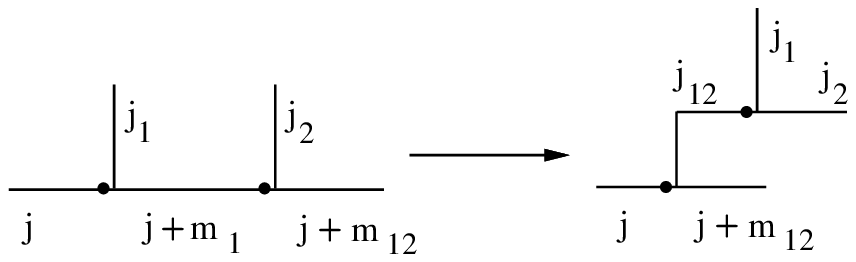}
\label{FusOneCond}
\eeq
\beq
\epsffile{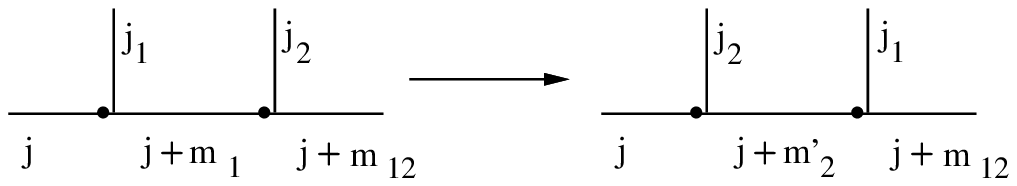}
\label{BrdOneCond}
\eeq
There are several equivalent forms, but the most convenient one, from the
viewpoint of polynomial equations is
$$
\sixjxi j_1,j_3,j_2,j_{123},j_{12},j_{23} _q =
{
\Xi _{j_2j_3}^{j_{23}}\
\Xi_{{j_1}j_{23}}^{j_{123}}
\over
\Xi_{j_1j_2}^{j_{12}}\
\Xi_{j_{12}j_3}^{j_{123}}
}
{
(-1)^{p_{12,3}+p_{1,2}}
\lfloor p_{2,3} \rfloor \! !
\lfloor p_{1,23} \rfloor \! !
\lfloor 2j_{23}+1 \rfloor
\over
\lfloor j_{23}+j_{123}-j_1+1 \rfloor _{p_{1,23}+p_{2,3}+1}
}
$$
$$
\!\!\!\!\!\!\!\!\!\!\!\!
\!\!\!\!\!\!\!\!\!\!\!\!
\sum_{\qquad\qquad
p_{1,23},p_{12,3}\le y\le p_{12,3}+p_{1,23},p_{12,3}+p_{1,2}}
\!\!\!\!\!\!\!\!\!\!\!\!
\!\!\!\!\!\!\!\!\!\!\!\!
\!\!\!\!\!\!\!\!\!\!\!\!
\!\!\!\!\!\!\!\!
(-1)^y
\lfloor j_1+j_{23}+j_{123}+2 \rfloor_{y-p_{1,23}}
$$
$$
{\lfloor y+2j_{123}-j_1-j_2-j_{12}+1 \rfloor_{p_{12,3}+p_{1,23}-y}
\over
\lfloor y-p_{1,23} \rfloor \! !
\lfloor y-p_{12,3} \rfloor \! !
}
$$
\beq
{
\lfloor y+2j_{123}-j_2-j_3-j_{23}+1 \rfloor_{p_{12,3}+p_{1,2}-y}
\lfloor j_2+j_{12}+j_{23}-j_{123}-y+1 \rfloor_{y-p_{12,3}}
\over
\lfloor p_{12,3}+p_{1,23}-y \rfloor \! !
\lfloor p_{12,3}+p_{1,2}-y \rfloor \! !
}
\label{sixj4}
\eeq
where the $p$'s are such that
$$
p_{1,2}\equiv j_1+j_2-j_{12},\quad
p_{2,3}\equiv j_2+j_3-j_{23},
$$
$$
p_{12,3}\equiv j_{12}+j_3-j_{123},\quad
p_{1,23}\equiv j_1+j_{23}-j_{123},
$$
so that
\beq
p_{k,l}\in Z_+,  \quad
p_{1,2}+p_{12,3}=p_{1,23}+p_{2,3}.
\label{p}
\eeq
For simplicity we lump  the square roots
of 6-j symbols into  coefficients noted  $\Xi$.
Those
$\Xi$ factors are chosen so that they can be
seen as normalization factors of the  vertices
and will cancel (or factorize) out
of the pentagonal equations when applying successive
fusions (or braidings).
So, the polynomial equations are fundamentally
rational equations (without square roots). We define
($p_{1,2}\equiv j_1+j_2-j_{12}$)
$$
\Xi_{j_1j_2}^{j_{12}}\equiv
\prod_{k=1}^{p_{1,2}}
\sqrt{
\lfloor 2j_1-k+1 \rfloor
\lfloor 2j_2-k+1 \rfloor
\lfloor 2j_{12}+k+1 \rfloor
\over
\lfloor k \rfloor
}
$$
\beq
=
\sqrt{
\lfloor j_1-j_2+j_{12}+1 \rfloor_{p_{1,2}}
\lfloor -j_1+j_2+j_{12}+1 \rfloor_{p_{1,2}}
\lfloor j_1+j_2+j_{12}+1 \rfloor_{p_{1,2}}
\over
\lfloor p_{1,2} \rfloor \! !
}
\label{Xi}
\eeq

We note that we only have the residual symmetry
\beq
\left\{ ^{j_1}_{j_3}\,^{j_2}_{j_{123}}
\right. \left |^{j_{12}}_{j_{23}}\right\}_q
=
\left\{ ^{j_3}_{j_1}\,^{j_2}_{j_{123}}
\right. \left |^{j_{23}}_{j_{12}}\right\}_q
\label{symsixj}
\eeq
The other symmetries are lost due to the particular choice
of the quantities $p_{k,l}$ to be positive integers.

One may show that all the polynomial equations are obeyed by
our generalization. thus we conclude that the fusion and braiding
are still given by Eqs.\ref{2.52}, \ref{2.57}, where all symbols
are replaced by the generalized ones.
\section{The case of two screening charges}
Since this case is somewhat tedious, partly due to notational
complications, we only give the results. The quantum group structure
is now $U_q(sl(2)) \odot U_{\qhat}(sl(2)$, with $q=\exp(ih)$,
and $\qhat =\exp(\qhat)$, and $h$ and $\hhat$ are given by Eq.\ref{a4.28}.
It describes the fusion and braiding of the general operators
$V_{m \mhat}^{(J \Jhat)}$ generated by the fusion and braiding of
$V_{m 0}^{(J 0)}=V_{m }^{(J)}$, and
$V_{0\mhat}^{(0 \Jhat)}=V_{\mhat}^{(\Jhat)}$. For half integer spins, the
$\odot$ product is a sort of graded tensor product, since
$V_{m }^{(J)}$ and $V_{\mhat}^{(\Jhat)}$ commute up to a
phase. This is not true for continuous $J$ where a completely novel
structure emerges. The fusion and braiding of
the general chiral operators $V_{\mge}^{(\Jge )}$, also denoted
$V_{m \mhat}^{(J \Jhat)}$, where underlined  symbols denote double indices
$\Jge \equiv (J,  \, \Jhat)$, $\mge \equiv (m,\,  \mhat)$,
takes the form
$$
{\cal P}_{\Jgen{} } V_{\Jgen23 -\Jgen{} }
^{(\Jgen1 )}(z_1)
V_{\Jgen3 -\Jgen23 }^{(\Jgen2 )}(z_2) =
\sum_{\Jgen12 }
{g_{\Jgen1 \Jgen2 }^{\Jgen{12} }\
g_{\Jgen{12} \Jgen3 }^{\Jgen{} }
\over
g _{\Jgen2 \Jgen3 }^{\Jgen{23} }\
g_{{\Jgen1 }\Jgen{23} }^{\Jgen{} }
}
\left\{
^{\Jgen1 }_{\Jgen3 }\,^{\Jgen2 }_{\Jgen{} }
\right. \left |^{\Jgen{12} }_{\Jgen{23} }\right\}_q \times
$$
\beq
{\cal P}_{\Jgen{} } \sum_{\{\nu\}}
V_{\Jgen3 -\Jgen{} } ^{(\Jgen12 ,\{\nu\} )}(z_2)
<\!\varpi _\Jgen12 ,\{\nu\}  \vert
V ^{(\Jgen1 )}_{\Jgen2 -\Jgen12 } \vert \varpi_{\Jgen2 }(z_1-z_2)\! >.
\label{2.1}
\eeq
$$
{\cal P}_{\Jgen{} } V_{\Jgen23 -\Jgen{} }
^{(\Jgen1 )}(z_1)
V_{\Jgen3 -\Jgen23 }^{(\Jgen2 )}(z_2) =
\sum_{\Jgen13 }
e^{\pm i\pi (\Delta_\Jgen{} +\Delta_\Jgen3
-\Delta_{\Jgen23 }-\Delta_{\Jgen13 })}\times
$$
\beq
{g_{\Jgen1 \Jgen3 }^{\Jgen{13} }\
g_{\Jgen{13} \Jgen3 }^{\Jgen{} }
\over
g _{\Jgen2 \Jgen3 }^{\Jgen{23} }\
g_{{\Jgen1 }\Jgen{23} }^{\Jgen{} }
}
\left\{
^{\Jgen1 }_{\Jgen2 }\,^{\Jgen3 }_{\Jgen{} }
\right. \left |^{\Jgen{12} }_{\Jgen{23} }\right\}_q
{\cal P}_{\Jgen{} }
V_{\Jgen13 -\Jgen{} } ^{(\Jgen2 )}(z_2)
V_{\Jgen3 -\Jgen13 }^{(\Jgen1 )}(z_1),
\label{c2.2}
\eeq
In these  formulae, the symbol $\varpi_{\Jge}$ stands for
$\varpi_0+2J+2\Jhat \pi/h$ where $\varpi_0 =1+\pi/h$
corresponds to the $sl(2)$-invariant
vacuum; ${\cal P}_{\Jgen{} }$ is the projector on
${\cal H}(\varpi_{\Jge})$. The above formulae contain the recoupling
coefficients (q-6-J symbols)  for the
quantum group structure $U_q(sl(2))\odot U_\qhat(sl(2))$.
For half integer spins,  they  are defined by
\beq
\left\{
^{\Jgen1 }_{\Jgen3 }\,^{\Jgen2 }_{\Jgen{} }
\right. \left |^{\Jgen{12} }_{\Jgen{23} }\right\}_q
=
(-1)^{ \fusV 1,2,,23,3,12 }
\left\{
^{J_1}_{J_3}\,^{J_2}_{J}
\right. \left |^{J_{12}}_{J_{23}}\right\}_q
\gaghat
\,^{\Jhat_1}_{\Jhat_3}\,^{\Jhat_2}_{\Jhat}
\bigr. \bverthat \, ^{\Jhat_{12}}_{\Jhat_{23}}\gadhat_{\qhat}
\label{x2.3}
\eeq
where $\fusV 1,2,,23,3,12 $ is an integer   given by
$$
\fusV 1,2,123,23,3,12
=
$$
\beq
2\Jhat_2(J_{12}+J_{23}-J_2-J_{123})
+2J_2(\Jhat_{12}+\Jhat_{23}-\Jhat_2-\Jhat_{123})
\label{sign}
\eeq
This sign displays the grading associated with the $\odot$ product
for the half integer case.
The symbol  $\left\{
^{J_1}_{J_3}\,^{J_2}_{J}
\right. \left |^{J_{12}}_{J_{23}}\right\}_q $ is the 6j coefficient
associated with  $U_q(sl(2))$, while $\gaghat
\,^{\Jhat_1}_{\Jhat_3}\,^{\Jhat_2}_{\Jhat}
\bigr. \bverthat \, ^{\Jhat_{12}}_{\Jhat_{23}}\gadhat_{\qhat}$
stands for the
q-6j  associated with $U_{\qhat}(sl(2))$. In addition to these group
theoretic features  there appear the coupling constants
$g_{{\Jgen{1} }\Jgen{2} }^{\Jgen{12} }$ whose general expression
is given in ref\cite{GS1}.

For continuous spins, the  $J$ $\Jhat$  quantum numbers loose meaning and
only the sum $\Je\equiv J+\Jhat\pi/h$
is significant. Concerning the states, a similar phenomenon occurs.
For  a state associated with   half-integer spins
$J$ and $\Jhat$, the corresponding zero-mode is
$\varpi=\varpi_{J,\Jhat}\equiv \varpi_0+2J+2\Jhat\pi/h$.
It is only a function of  $\Je$
and  we have $\varpi_{J,\Jhat}=\varpi_0+2\Je$, also
denoted $\varpi_{\Je}$.
One may verify that
 the fusion and braiding matrices may be written  in
terms of these  effective spins.
Of course, if $h$ is irrational, using $J$, $\Jhat$ or $\Je$
is immaterial. In the  half-integer spin case,
there is a remarkable fact, which is related:
by using the properties of the gamma functions under integer shifts
and of the sinus functions under shifts by ($\pi\times$integer) one
may actually absorb the sign factors into the 6-j's. It is this
last form which carries over to the continuous case, where one has,
for instance,
\beq
F_{{\Je_{23}},{\Je_{12}}}\!\!\left[^{\Je_1}_{\Je_{123}}
\,^{\Je_2}_{\Je_3}\right]
=
{g_{\Je_1\Je_2}^{\Je_{12}}\
g_{\Je_{12}\Je_3}^{\Je_{123}}
\over
g _{\Je_2\Je_3}^{\Je_{23}}\
g_{{\Je_1}\Je_{23}}^{\Je_{123}}
}
\left\{\left\{ ^{\Je_1}_{\Je_3}\,^{\Je_2}_{\Je_{123}}
\left. \right |^{\Je_{12}}_{\Je_{23}}\right\}\right\}
\gaghat\gaghat\, ^{\Jehat_1}_{\Jehat_3}\,^{\Jehat_2}_{\Jehat_{123}}
\bverthat\bigr. \, ^{\Jehat_{12}}_{\Jehat_{23}}\gadhat\gadhat.
\label{Fjjhat}
\eeq
the 6-j symbols with double braces are ordinary symbols with shifted
arguments:
\beq
\left\{
\left\{ ^{\Je_1}_{\Je_3}\,^{\Je_2}_{\Je_{123}}
\right. \left |^{\Je_{12}}_{\Je_{23}}\right\}\right\}
\equiv
\left\{ ^{\Je_1}_{\Je_3}\,^{\Je_2}_{\Je_{123}+(\phat_{2,3}+\phat_{1,23})\pi/h}
\right. \left |^{\Je_{12}+\phat_{1,2}\pi/h}_
{\Je_{23}+\phat_{2,3}\pi/h}\right\}_q,
\label{sixjp}
\eeq
with similar relations for the hatted ones. It is only for half integer
$J$ that one may disentangle the hatted and unhatted quantum numbers.
\section{Applications}
As is well know there are two completely different regimes. The explicit
formulae for the quantum group parameters are
$$
h={\pi \over 12}\Bigl(C-13 -
\sqrt {(C-25)(C-1)}\Bigr),
$$
\beq
\hhat={\pi \over 12}\Bigl(C-13
+\sqrt {(C-25)(C-1)}\Bigr),
\label{51.1}
\eeq
The weak coupling regime ($C\geq 25$, and $C\leq 1$) is such that
$h$ and $\hhat$ are real. In the strong coupling one ($1\leq C \leq 25$),
on the contrary, $h$ and $\hhat$ are complex (conjugate).
\subsection{The weak coupling regime}
Here the point is to reconstruct the Liouville exponentials so that
locality is ensured.
The orthogonality relations Eqs.\ref{x3.25}
are precisely what we need for this purpose.
 Indeed we shall
write,
for arbitray $J$,
\beq
e^{\textstyle -J\alpha_-\Phi(\sigma, \tau )}=
\sum _{m=-J}^{\infty}
\Vt_m^{(J)}(z)\,
{\overline \Vt_{m}^{(J)}}(\zb) \>
\label{53.27}
\eeq
 We assume that
$\Vt_m^{(J)}(z)$, and ${\overline \Vt_{m}^{(J)}}(\zb)$ commute.
According to Eq.\ref{3.8}, the braiding of the $\Vt$ fields is equal to the
6-j symbols.
  Thus the braiding
matrix of the $\Vt$ fields is the same as the one of the $\overline \Vt$
fields, and we have
$$
e^{\textstyle -J_1\alpha_-\Phi(\sigma_1, \tau )}
e^{\textstyle -J_2\alpha_-\Phi(\sigma_2, \tau )}=
$$
$$
\sum_{m_1, m_2}
\sum_{n_1, n_2; \nb_1, \nb_2}
\left\{ ^{J_1\quad }_{J_2\quad }
\> ^{( \varpi-\varpi_0 +2m_1+2m_2)/2}
_{ \quad \quad (\varpi-\varpi_0)/ 2}
\right.
\left |^{\quad ( \varpi-\varpi_0 +2n_2)/2}
_{\quad ( \varpi-\varpi_0+2m_1)/2}\right\}_q
$$
$$
\times
\left\{ ^{J_1\quad }_{J_2\quad }
\> ^{( \varpib-\varpi_0 +2m_1+2m_2)/2}
_{ \quad \quad (\varpib-\varpi_0)/ 2}
\right.
\left |^{\quad ( \varpib-\varpi_0 +2\nb_2)/2}
_{\quad ( \varpib-\varpib_0+2m_1)/2}\right\}_q
$$
\beq
\times
\left. \Vt_{n_2}^{(J_2)}(z_2) \Vt_{n_1}^{(J_1)}(z_1)\,
{\overline \Vt_{n_2}^{(J_2)}}(\zb_2) {\overline \Vt_{n_1}^{(J_1)}}(\zb_1)
 \right |_{n_1+n_2=m_1+m_2 \atop
\nb_1+\nb2=m_1+m_2}.
\label{53.28}
\eeq
We only discuss the case where $\varpi=\varpib$ (no winding number)
for simplicity. Then, the summation over $m_1$ in the last equation
precisely coincides with the summation over $J_{23}$ in Eq.\ref{x3.25}.
This gives immediately
\beq
e^{\textstyle -J_1\alpha_-\Phi(\sigma_1, \tau )}
e^{\textstyle -J_2\alpha_-\Phi(\sigma_2, \tau )}=
e^{\textstyle -J_2\alpha_-\Phi(\sigma_2, \tau )}
e^{\textstyle -J_1\alpha_-\Phi(\sigma_1, \tau )},
\label{53.29}
\eeq
and the Liouville exponential is local for arbitrary $J$.
Since we have defined the Liouville exponential for countinuous $J$ we may
obtain the Liouville field itself by computing
\beq
\Phi(\sigma, \tau )\equiv -{1\over \alpha_-}
 \left. {d e^{\textstyle -J\alpha_-\Phi(\sigma, \tau )}
\over dJ}\right |_{J=0}
\label{53.30}
\eeq
One may verify\cite{GS3}  that the canonical commutation relations  hold:
\beq
\left [ \Phi(\sigma, \tau ) \dot \Phi(\sigma', \tau )\right ]=
4\pi i \gamma \delta(\sigma-\sigma'),
\label{53.31}
\eeq
and that the quantum Liouville equation is satisfied.

So far the only direct application of the weak coupling formulae
is the derivation of the matrix-model results on the sphere
given in ref.\cite{G5}. One couples the Liouville theory with
another copy of the same theory with central
charge $c=26-C$,  which represents matter. For $C>25$, we have
$c<1$.
The resulting three-point function on the sphere
is a product of leg factors as expected.

\subsection{The strong coupling regime}
The problem is the so called ``$c=1$ barrier''. The present
approach is the only way\cite{GN,GR1,G3} to go through it.
The basic reason seems to
be that one
should treat the two screening charges symmetrically in the strong coupling
regime, since they are complex conjugate. This is in sharp
contrast with what is
currently done in the weak coupling regime, say using matrix models.
In the strong coupling regime $1\leq C \leq 25$,
$h$ and $\hhat$ are complex conjugate.
Thus,  treating them symmetrically, as was done in refs.\cite{GN,GR1,G3},
 is the key.
The basic result is the derivation of truncation theorems that hold
for special
values of $C$. The point of these theorems is as follows.
The basic family of $V_{m\mhat}^{(J \Jhat)}$ operators
has
Virasoro weights  given by
\beq
 \Delta_{J\Jhat}
={C-1\over 24}-
{ 1 \over 24} \left((J+\Jhat+1) \sqrt{C-1}
-(J-\Jhat) \sqrt{C-25} \right)^2,
\label{51.2}
\eeq
in agreement with Kac's formula.
In the strong coupling regime $\sqrt{C-1}$ is real, and $\sqrt{C-25}$
pure imaginary so that the  formula just written
 give complex results in general.
However---in a way that is reminiscent
of the truncations that give the minimal
unitary models--- for $C=7$,
$13$, and $19$,  there is a consistent
truncation of the above general family
down to an operator algebra involving
operators with real Virasoro conformal
weights only. These  are of two types.
The first has spins $\Jhat=J$, and Virasoro weights that are negative;
the second has $\Jhat=-J-1$, and  Virasoro weights that are
positive.
the truncation theorems make use of the following notions.

\smallskip

\noindent \souligne{a) The physical Hilbert spaces.} They are of the
form
\beq
{\cal H}^\pm_{s \,  \hbox {\scriptsize phys}}
\equiv  \bigoplus_{r=0}^{1\mp s}\,\bigoplus_{n=-\infty}^{_\infty}\>
{\cal H}_s\bigl(\varpi^\pm _{r,\, n}\bigr),
\label{52.16}
\eeq
\beq
\varpi^\pm_{r,\, n}
\equiv \Bigl({r\over 2\mp s}+n\Bigr)
\,\bigl(1\mp {\pi\over h}\bigr).
\label{52.17}
\eeq
 The integer $s$ is such that the special values correspond to
\beq
 C=1+6(s+2),\quad s=0,\,\pm 1;\quad h+\hhat=s \pi.
\label{52.7}
\eeq
For the plus sign, the weight
$\Delta (\varpi^+_{r,\, n})
\equiv (1+\pi/h)^2h/4\pi -h \varpi_{r,\, n}^2/4\pi$ is positive and
 in
${\cal H}^+_{s \,  \hbox {\scriptsize phys}}$,
 the representation of the Virasoro algebra
is
  unitary. For the minus sign $\Delta (\varpi^-_{r,\, n})$ is real but
negative.  This latter case is only useful for topogical theories.
In ${\cal H}^+_{s \,  \hbox {\scriptsize phys}}$ the  partition
function corresponds to compactification on a circle
with radius $R=\sqrt {2(2-s)}$ (see refs.\cite{GR1}).
\smallskip

\noindent\souligne{b) The restricted set of conformal weights}. The
truncated family only involves
 operators of the type   $(2J+1,2J+1)$ noted $\chi_-^{(J)}$
and $(-2J-1, 2J+1)$ noted $\chi_+^{(J)}$. Their Virasoro
conformal weights\cite{GN,G3,GR1} which are
respectively given by
\beq
\Delta^-(J, C)=-{C-1\over 6}\,J(J+1), \quad
\Delta^+(J, C)=1+{25-C\over 6}\, J(J+1),
\label{52.8}
\eeq
are real.  $\Delta^-(J)$ in negative for all $J$ (except for
$J=-1/2$
where it becomes equal to $\Delta^+(-1/2)=(s+2)/4$). $\Delta^+(J)$
is always positive, and is larger than one if $J \not= -1/2$.\smallskip

\noindent\souligne{c) The truncated families}:
${\cal A}^\pm_{phys}$
is the   set of operators noted $\chi_\pm^{(J)}$,
introduced in \cite{GN,G3,GR1}, whose conformal weights are given by
Eq.\ref{52.8}. they are of the form
\beq
{\cal P}_{\Jme_{12}}\chi_-^{(J_1)}
\equiv
\sum_{J_2,p_{1,2}\in Z_+}
(-1)^{(2+s)(2J_2p_{1,2}
+{p_{1,2}(p_{1,2}+1)\over 2})}
g_{\Jme_1\Jme_2}^{\Jme_{12}}
{\cal P}_{\Jme_{12}}
V^{(\Jme_1)}
{\cal P}_{\Jme_2}
\label{chi-}
\eeq
\beq
{\cal P}_{\Jpe_{12}}\chi_+^{(J_1)}
\equiv
\sum_{J_2,p_{1,2}\in Z_+}
(-1)^{(2-s)(2J_2p_{1,2}
+{p_{1,2}(p_{1,2}+1)\over 2})}
g_{\Jpe_1\Jpe_2}^{\Jpe_{12}}
{\cal P}_{\Jpe_{12}}
V^{(\Jpe_1)}
{\cal P}_{\Jpe_2}
\label{chidef+}.
\eeq
This definition is such that the following holds

\smallskip

\noindent {THE TRUNCATION THEOREMS}:

For $C=1+6(s+2)$, $s=0$, $\pm 1$,
and when it
acts on ${\cal H}^+_{s \,  \hbox {\scriptsize phys}}$
(resp. ${\cal H}^-_{s \,  \hbox {\scriptsize phys}}$);
 the  set ${\cal A}^+_{phys}$
(resp. ${\cal A}^-_{phys}$)
of operators $\chi_{+}^{(J)}$ (resp. $\chi_{-}^{(J)}$)
 is closed by fusion and braiding, and
  only gives states that belong to
${\cal H}^+_{s \,  \hbox {\scriptsize phys}}$
(resp. ${\cal H}^-_{s \,  \hbox {\scriptsize phys}})$.

\smallskip
Let us now turn to the new topological models put
forward in ref.\cite{G3}. One considers two copies of
the above strongly coupled theories with central charges
$C=1+6(s+2)$, and $c=1+6(-s+2)$. They   play the
role of gravity and matter respectively. This gives
consistent theories since clearly $C+c=26$.
As shown in ref.\cite{GR1}, the three-point functions are
calculable and given by a product of $g$ factors of the type
Eq.\ref{3.7}. The result is again a product of leg
factors, and the higher point functions seem calculable
using methods similar to the one used for the weak coupling regime.
The novel feature is that now the vertex
operators are  expressed in terms of the chi fields
given by Eqs.\ref{chi-} \ref{chidef+} --- and their antiholomorphic
counterparts --- instead of the
Liouville exponentials. As a result they do not preserve any longer
the equality between left- and right-quantum group spins.
Thus the transition through the $c=1$ barrier seems to be
characterized by a sort of deconfinement of chirality.

\end{document}